\documentclass{article}
\usepackage{amsfonts}
\usepackage{amsmath}
\usepackage{amssymb}
\usepackage{float}
\usepackage{geometry}
\usepackage{graphicx}
\usepackage{epstopdf}
\usepackage{bm}
\usepackage{stackrel}
\usepackage[T1]{fontenc}
\usepackage[utf8]{inputenc}
\usepackage{authblk}
\usepackage{ulem}
\usepackage{cancel}
\usepackage[pdfstartview=FitH, CJKbookmarks=true,
bookmarksnumbered=true, bookmarksopen=true,
colorlinks,
linkcolor=green,
anchorcolor=blue, citecolor=blue
]{hyperref}

\newcommand{\comment}[1]{}
\newcommand{\EEA}{\end{eqnarray}}
\newcommand{\BEA}{\begin{eqnarray}}
\newcommand{\peq}{p_{eq}}
\providecommand{\keywords}[1]{\textbf{{Keywords.}} #1}
\providecommand{\AMS}[1]{\textbf{{Mathematics Subject Classification.}} #1}

\begin{document}
\title{Modeling of Missing Dynamical Systems: Deriving Parametric Models using a Nonparametric Framework}
%\author{S.W. Jiang and J. Harlim}
\author[a]{Shixiao W. Jiang }
\author[a,b,c]{John Harlim \thanks{Corresponding author: jharlim@psu.edu} }
%\cortext[cor]{Corresponding author}
\affil[a]{Department of Mathematics, the Pennsylvania State University, 109 McAllister Building, University Park, PA 16802-6400, USA}
\affil[b]{Department of Meteorology and Atmospheric Science, the Pennsylvania State University, 503 Walker
Building, University Park, PA 16802-5013, USA}
\affil[c]{Institute for Computational and Data Sciences, the Pennsylvania State University, 224B Computer Building, University Park, PA 16802, USA}
\date{\today}
\maketitle

\begin{abstract}
In this paper, we consider modeling missing dynamics with a nonparametric non-Markovian model, constructed using the theory of kernel embedding of conditional distributions on appropriate Reproducing Kernel Hilbert Spaces (RKHS), equipped with orthonormal basis functions. Depending on the choice of the basis functions, the resulting closure model from this nonparametric modeling formulation is in the form of parametric model. This suggests that the success of various parametric modeling approaches that were proposed in various domains of applications can be understood through the RKHS representations.
When the missing dynamical terms evolve faster than the relevant observable of interest, the proposed approach is consistent with the effective dynamics derived from the classical averaging theory. In the linear Gaussian case without the time-scale gap, we will show that the proposed non-Markovian model with a very long memory yields an accurate estimation of the nontrivial autocovariance function for the relevant variable of the full dynamics. Supporting numerical results on instructive nonlinear dynamics show that the proposed approach is able to replicate high-dimensional missing dynamical terms on problems with and without the separation of temporal scales.
\end{abstract}

\


\

\AMS{37M10,37M25,41A10,41A45,60J22}
%\begin{keywords}
%Something \sep%
%    Something else
%\end{keywords}


\section{\label{sec:intro} Introduction}

One of the long-standing issues in modeling dynamical systems is model errors arising from incomplete understanding of the physics. The progress in tackling this problem goes under different names depending on the scientific fields. In applied mathematics and engineering sciences, some of these approaches are known as the {\it reduced-order modeling}, which ultimate goal is to derive an effective model with low computational complexity from the first principle, assuming that the full dynamics is known. They include the Mori-Zwanzig formalism \cite{Mori65,Zwanzig61,Zwanzig73} and its approximations \cite{chk:02,chorin2007problem,gouasmi2017priori,hl:15,kondrashov2015data}; the averaging/homogenization when there is an apparent scale separation between the relevant and irrelevant variables \cite{givon2004extracting,majda1999models,majda2001mathematical,eweinan2007heterogeneous}. In domain sciences, various methods for {\it subgrid-scale parameterization} were proposed to handle the same problem that arises in applications such as material science, molecular dynamics, climate dynamics, just to name a few. They include the Markov chain type modeling \cite{crommelin2008subgrid,kbm:10}; stochastic parameterization \cite{bh:14,crommelin2008subgrid,hmm:14,kwasniok2012data,lu2016comparison,lu2017accounting,mh:13,wilks2005effects}; superparameterization in cloud modeling \cite{grabowski:04,kscmt:11,mg:14} and in combustion problems \cite{kerstein:88,kerstein:99}; Direct-Interaction Approximation (DIA) for parameterizing sub-grid scale processes in isotropic turbulence \cite{kraichnan:59} and its extensions \cite{fo:08}, for modeling non-Markovian memory in inhomogeneous turbulence over topography. We should point out that this list is incomplete and these approaches share some commonality despite being developed independently and having different implementation details. Namely, the key unifying theme in these aforementioned methods is the parametric modeling assumption with specific choices of class of functions/distributions and typically having finite number of parameters.

In this paper, we consider a nonparametric modeling framework to compensate for the missing dynamical components. {\color{black}One of the main goals in this paper is to show that parametric modeling approaches can be understood and systematically derived from a nonparametric framework as opposed to the empirical choices of parametric models.}
In our setup, suppose that the underlying full dynamics is an ergodic system of It\^{o} diffusion with relevant components $x\in \mathcal{X}$ and irrelevant components $y\in\mathcal{Y}$. The objective is to predict the evolution of $x\in\mathcal{X}$ and its statistics, given only the $x-$component of the full dynamics,
\begin{equation}
\begin{aligned}\label{dynamics}
dx &= a(x,y)\,dt + b(x,y)\,dW_t, \\
dy &= c(x,y)\,dt + d(x,y)\,dV_t,
\end{aligned}
\end{equation}
and a historical data set $\{x_i:=x(t_i),y_i=y(t_i)\}_{i=1,\ldots,N}$. In \eqref{dynamics}, $a$ and $b$ denote, respectively, the $x-$component of the drift and diffusion terms that are known, while $c$ and $d$ denote, respectively, the $y-$component of the drift and diffusion terms that are not known. Also, $W_t$ and $V_t$ denote the standard (uncorrelated) Wiener processes.

{\color{black}While the core of the problem is similar to that in the reduced-order modeling framework, the fact that we have no knowledge of the full dynamics prohibits us to derive an effective equation from the first principle as in the standard averaging theory or Mori-Zwanzig formalism.} Motivated by the practical applications where the underlying dynamics are not fully understood, instead, we will use the available historical data to reconstruct the missing dynamical components. We should point out that the restriction of knowing historical measurement of the irrelevant component, $y_i \in\mathcal{Y}$, can be relaxed in some cases. When $\{x_i \}_{i=1\ldots,N}$ is the only available measurement, one can use, for example, likelihood maximum estimate \cite{kkg:05,lu2017data} in the deterministic case or an adaptive Bayesian filtering \cite{bh:16jcp} (when $b$ is constant and the training data is noisy) to extract the ``identifiable'' components of $y_i$. By identifiable components, we refer to variables that depend on $y$ that appear in $a$ and $b$, as we shall see in our numerical examples. Abusing the notation, we will denote $y$ as the identifiable components. We will clarify this notion in our numerical examples.

Given the pair of historical time series $\{x_i,y_i\}_{i=1,\ldots,N}$ with time lag $\tau=t_{i+1}-t_i$, let us define $\bm{z}_t := (\bm{x}_{t-m:t},\bm{y}_{t-n:t-1})\in\mathcal{Z}$ with $\bm{x}_{t-m:t}:=(x_{t-m},x_{t-m+1},\ldots,x_t)$ and $\bm{y}_{t-n:t-1}:=(y_{t-n},y_{t-m+1},\ldots,y_{t-1})$ for some integers $m, n \in\{-1,0,\ldots\}$. When $m=-1$, $\bm{z}_t$ has only $\bm{y}$ components (similarly for $n=0$, $\bm{z}_t$ has only $\bm{x}$ components). {\color{black}We should point out that we have reserved the index $i$ for the training data and used a different index $t$ for an arbitrary prediction time with the same lag $\tau$.
Given these time series, our modeling approach is to approximate the conditional expectations,
\BEA
\hat{a}(x,\bm{z}_t) := \mathbb{E}[a(x,Y)|\bm{z}_t] \quad\quad \hat{B}(x,\bm{z}_t) := \mathbb{E}[b(x,Y)b(x,Y)^\top|\bm{z}_t], \label{conditionalestimator}
\EEA
where the expectations are defined with respect to the equilibrium conditional density $p(y|\bm{z}_t)$ of the random variable $Y|\bm{z}_t$, a short hand for $Y|\bm{z}=\bm{z}_t$. {\color{black} Here, $Y|\bm{z}$ is nothing but the stationary random variable $Y_t|\bm{z}_t$. Throughout this paper, we will not use the notation $Y_t|\bm{z}_t$ to avoid a potential confusion with the non-stationary time-dependent distributions.}

Given these conditional statistics, the closure model is given by
\BEA
\hat{x}_{t+1} =  \hat{x}_{t} + \int_{t\tau}^{(t+1)\tau} \hat{a}(\hat{x}(s),\bm{\hat{z}}_t)\, ds + \int_{t\tau}^{(t+1)\tau} \hat{B}(\hat{x}(s),\bm{\hat{z}}_t)^{1/2} dW_s,\label{abstractrom}
\EEA
where $\bm{\hat{z}}_t := (\bm{\hat{x}}_{t-m:t},\bm{\hat{y}}_{t-n:t-1})$. To proceed the forecast at the next time step-$(t+1)$, one needs to update $\bm{\hat{z}}_{t+1}$. This variable is obtained by concatenating the components from previous time steps, $(\bm{\hat{x}}_{t+1-m:t+1},\bm{\hat{y}}_{t+1-n:t-1})$ and $\hat{y}_{t}=\mathbb{E}[Y|\bm{\hat{z}}_t]$ that is estimated at time $t$.}

\comment{
{\color{red}To implement this model in iterative mode, one also concatenates the known components for $\bm{\hat{z}}_{t+1}$, and then updates $\hat{y}_{t+1}\cancel{=\mathbb{E}[Y|\bm{\hat{z}}_{t+1}]}$ at each time step as follows:  }
{\color{red}
\begin{equation}
\begin{aligned}\label{mode_append}
\bm{\hat{z}}_{t+1}&=(\bm{\hat{x}}_{t+1-m:t+1},\bm{\hat{y}}_{t+1-n:t}),\\
\hat{y}_{t+1}&=\mathbb{E}[Y|\bm{\hat{z}}_{t+1}],
\end{aligned}
\end{equation}
}}
Notice that if the $x-$component is slow and the missing $y-$component is fast with a scale gap denoted by a small parameter $\epsilon$, the closure model in \eqref{abstractrom} is identical to the effective dynamics deduced by the averaging theory \cite{khasminskii:68,kurtz:75,papanicolaou1976some} when the conditional expectation in \eqref{conditionalestimator} is defined with respect to the invariant density of the fast dynamics $\rho_\infty(y;\hat{x}_t)$ for a fixed $\hat{x}_t$, if such density exists. In this specific situation (fast-slow system), by setting $\bm{\hat{z}}_t=\hat{x}_t$, that is $m=0, n=0$, our framework effectively closes the dynamics by averaging over $p(y|\hat{x}_t) = p_{eq}(\hat{x}_t,y)/\int p_{eq}(\hat{x}_t,y)dy$, where $p_{eq}$ denotes the invariant density of the full dynamics. We will show that averaging over $p(y|\hat{x}_t)$ is consistent with averaging over $\rho_\infty (y;\hat{x}_t)$ up to order $\epsilon$.  In general case where there is no separation of scales, the choice of $m,n$ will be problem dependent.
%When $m$ and/or $n$ are strictly positive, we refer to the conditional density $p(y_{t}|\bm{z}_t)$ as a \emph{non-Markovian transition density}.
In this case, the predictive skill of certain statistics will depend on the specific choices of $\bm{z}_t$.  For example, in the linear Gaussian case without a time-scale gap, we will show the existence of a conditional density $p(y|\bm{z})$ which allows for Eq.~\eqref{abstractrom} to accurately estimate one-point and two-point statistics of the $x$-components of the full dynamics.

%Numerically, the integration over the phase space $\mathcal{Y}$ will be approximated via Monte-Carlo average over the training data set. The time integration will be realized with appropriate SDE's solvers.

The main idea in this paper is to consider a nonparametric representation for $p(y|\bm{z})$ using the theory of kernel embedding of conditional distributions, which was introduced in the machine learning community \cite{Song2013IEEE,Song2009hilbert}. {\color{black}The kernel embedding of conditional distributions \cite{Song2013IEEE,Song2009hilbert} suggests that one can represent probability distribution as an element of a reproducing kernel Hilbert space (RKHS). In this paper, we will show that if $\mathcal{H}$ is an RKHS induced by an orthonormal basis $\{\phi_k:\mathcal{Y}\to\mathbb{R}\}$ of an appropriate $L^2-$space, then any $p(\cdot| \bm{z})\in\mathcal{H}$ for any fixed $\bm{z}\in \mathcal{Z}$ can be represented as $p(\cdot| \bm{z}) = \sum_{k=1}^\infty c_k(\bm{z}) \phi_k(\cdot)$, where the coefficients in $c_k$ will be pre-computed using the historical data set and the kernel embedding of conditional distributions formula. Here, the convergence of the series representation is in uniform sense.} In this paper, we will consider parametric orthonormal basis functions such as the Hermite polynomials for low-dimensional $\mathcal{Z}$ as well as the proper orthogonal decomposition (POD) modes for high-dimensional $\mathcal{Z}$. In the latter case, we shall see that the resulting closure model in \eqref{abstractrom} is a parametric model that is well-known, namely the linear non-autonomous autoregressive model. In general, the form of parametric closure models depends on the ansatz of $\phi_k$ as a function of $\bm{z}$. We should point out that one can also leave it entirely nonparametric with the data-driven basis functions constructed by the diffusion maps algorithm as in \cite{Berry2017MWR,jiang2018parameter}. While this is theoretically sound, the construction of data-driven basis functions requires an elaborate computational effort and is limited to problems with intrinsically low-dimensional $\mathcal{Y}$. In addition to constructing the basis, the main computational cost arises when evaluating the estimated basis functions on new points $\hat{z}_t$ for future-time prediction. Given these constraints, we will not explore the data-driven nonparametric basis in this paper.

The remaining of the paper is organized as follows. In Section~\ref{section2}, we  briefly review the theory of kernel embedding of conditional distributions for estimating $p(y|\bm{z})$ using an orthonormal basis representation and discuss the proposed closure models in detail.  In Section~\ref{section3}, we provide an intuition for choosing the density
$p(y|\bm{z})$ by discussing missing dynamics in a linear  Gaussian dynamics with and without temporal scale gaps. In Section~\ref{sec:examples}, we numerically demonstrate the proposed approach on two nonlinear high-dimensional test problems, {\color{black}where $m, n$ are small in the first example and large in the second example.} In Section~\ref{section5}, we conclude the paper with a brief summary and discussion. {\color{black}We supplement the paper with two Appendices: Appendix~\ref{app:A} supplements Section~\ref{section2} with a more detailed derivation of the kernel embedding of conditional distributions; Appendix~\ref{app:ACF_linear} shows the consistency of the proposed approach in estimating autocovariance functions in the linear Gaussian case without the time-scale gap.}

\section{A nonparametric formulation of modeling missing dynamics}\label{section2}

In this section, we first give a brief review on the kernel embedding of conditional distributions introduced in \cite{Song2013IEEE,Song2009hilbert}, formulated using an orthonormal basis of appropriate square-integrable function spaces as in \cite{Berry2017MWR,jiang2018parameter,ZHL:19}. Subsequently, we present the proposed nonparametric modeling approach for missing dynamics. % and discuss the consistency for problems with and without scale gaps.

\subsection{Kernel embedding of conditional distributions}\label{sec21}

Let $\mathcal{Y}$ be a compact set and define $K:\mathcal{Y}\times\mathcal{Y}\to\mathbb{R}$ to be a kernel, which means it is symmetric positive definite and let it be bounded. By Moore-Aronszajn theorem, there exists a unique Hilbert space $\mathcal{H}=\overline{\mbox{span}\{K(y,\cdot),\forall y\in\mathcal{Y}\}}$. Let $q:\mathcal{Y}\to\mathbb{R}$ be a positive weight function and $\{\psi_k q\}_{k\geq 1}$ be a set of eigenfunctions corresponding to eigenvalues $\{\lambda_k\}$ of the following integral operator,
\BEA
\mathcal{K}f(y) = \int_{\mathcal{Y}} K(y,y')f(y')q^{-1}(y')dy', \quad \quad f\in L^2(\mathcal{Y},q^{-1}).\label{intoperator}
\EEA
We should note that $\{\psi_k q\}$ forms an orthonormal basis of $L^2(\mathcal{Y},q^{-1})$ and by Mercer's theorem, the kernel $K$ has the following representation,
\BEA
K(y,y') = \sum_{k=1}^{\infty} \lambda_k \psi_k(y)q(y)\psi_k(y')q(y').\label{mercer}
\EEA
We should point out that if $\mathcal{Y}$ is not a compact domain such as $\mathbb{R}^n$, with an exponentially decaying $q$, one can construct a bounded Mercer-type kernel as in \eqref{mercer} with an appropriate choice of decreasing sequence $\{\lambda_k\}$ (see Lemma 3.2 in \cite{ZHL:19}) {\color{black} and it is a reproducing kernel corresponding to the RKHS $\mathcal{H}$ (see Proposition 3.4 in \cite{ZHL:19}).} This result provides a justification for the use of Hermite polynomials $\{\psi_k\}$ with Gaussian weight $q$ in one of our numerical examples.

We should point out that the RKHS $\mathcal{H}$ induced by the Mercer-type kernel in \eqref{mercer} is a subspace of $L^2(\mathcal{Y},q^{-1})$ with the reproducing property corresponding to the inner product defined as $\langle f,g\rangle_\mathcal{H} = \sum_{k=1}^\infty \frac{f_k g_k}{\lambda_k}$, where $f_k = \langle f,\psi_kq\rangle_{L^2(\mathcal{Y},q^{-1})}$ and  $g_k = \langle g,\psi_kq\rangle_{L^2(\mathcal{Y},q^{-1})}$. Then for any $f\in \mathcal{H}$ and $y\in\mathcal{Y}$, we can represent,
\BEA
f(y) = \langle f,K(y,\cdot)\rangle_{\mathcal{H}} = \sum_{k=1}^\infty \frac{f_k \lambda_k \psi_k(y)q(y)}{\lambda_k} =  \sum_{k=1}^\infty f_k \psi_k(y)q(y),\nonumber
\EEA
with the orthonormal basis of $L^2(\mathcal{Y},q^{-1})$ where the convergence of the series holds uniformly (or in $C_0(\mathbb{R}^n)$ for non-compact $\mathcal{Y}=\mathbb{R}^n$).

\comment{Analogously, we define  $\mathcal{\hat{H}}$ to be an RKHS for functions of $\bm{z} \in\mathcal{Z}$, which can be represented by orthonormal basis $\Phi_l(\bm{z}):=\varphi_l(\bm{z})\hat{q}(\bm{z}) \in L^2(\mathcal{Z},\hat{q}^{-1})$.}

{\color{black}Let ${Y}$ and ${Z}$ be random variables on $\mathcal{Y}$ and $\mathcal{Z}$, respectively, with distribution $P(Y,Z)$. Assuming that the conditional density $p(\cdot|\bm{z})\in \mathcal{H}$ for any fixed $\bm{z}\in \mathcal{Z}$, we have the  representation for $p(\cdot|\bm{z})$ in the RKHS $\mathcal{H}$ of real-valued functions on $\mathcal{Y}$:
\BEA
p(y|\bm{z}) = \sum_{k=1}^{\infty} \langle p(\cdot|\bm{z}),\psi_kq\rangle_{L^2(\mathcal{Y},q^{-1})} \psi_k(y)q(y) =  \sum_{k=1}^{\infty} \langle p(\cdot|\bm{z}),\psi_k\rangle_{L^2(\mathcal{Y})} \psi_k(y)q(y),\label{conditionaldensity}
\EEA
where the convergence of the series is in the uniform sense and the coefficients are to be determined. The theory of kernel embedding of conditional distributions \cite{Song2013IEEE,Song2009hilbert}, implemented also in \cite{Berry2017MWR,jiang2018parameter}, suggests that the coefficients can be expressed as,
\BEA
\langle p(\cdot|\bm{z}),\psi_k\rangle_{L^2(\mathcal{Y})} = \mathbb{E}_{Y|\bm{z}}[\psi_k(Y)] = \sum_{l=1}^\infty \left[ \bm{C}_{%
{Y}{Z}}\bm{C}_{{ZZ}}^{-1}\right] _{kl}\varphi _{l}\left(
\bm{z}\right),\label{kme}
\EEA
where $\{\varphi_l\}_{l\geq 1}$ forms an orthonormal basis of $L^2(\mathcal{Z},\hat{q})$ and
\BEA
\left[\bm{C}_{{YZ}}\right] _{ks} =\mathbb{E}_{{YZ}}%
\left[ \psi _{k}(Y)\otimes \varphi _{s}(Z)\right], \quad\quad
\left[ \bm{C}_{{ZZ}}\right] _{sl} =\mathbb{E}_{{ZZ}}%
\left[ \varphi _{s}(Z)\otimes\varphi _{l}(Z)\right].\nonumber
\EEA
See the detailed derivation of \eqref{kme} in Appendix~\ref{app:A}. Substituting \eqref{kme} to \eqref{conditionaldensity}, we obtain,
\BEA
p( y| \bm{z}) =\sum_{k,l=1}^\infty\psi
_{k}\left(y\right)q(y) \left[ \bm{C}_{%
{Y}{Z}}\bm{C}_{{ZZ}}^{-1}\right] _{kl}\varphi _{l}\left(
\bm{z}\right). \label{Eqn:tranker}
\EEA
Notice that this representation can be understood as a linear regression in infinite-dimensional spaces with respect to the basis functions $\psi _{k}q$ and $\varphi_{l}$. Connecting to the notation in the introduction, $c_k(\bm{z})=\sum_{l=1}^\infty \left[ \bm{C}_{%
{Y}{Z}}\bm{C}_{{ZZ}}^{-1}\right] _{kl}\varphi _{l}\left(
\bm{z}\right)$ and $\phi_k=\psi_k q$. The representation in (\ref{Eqn:tranker}) is nonparametric in the sense that we do not assume any particular distribution for the density.
}

Given pairs of data $\{y_i,\bm{z}_i\}_{i=1,\ldots,N}$, where $\bm{z}_i := (\bm{x}_{i-m:i},\bm{y}_{i-n:i-1})\in\mathcal{Z}$, {\color{black} distributed according to $P(Y,Z)$}, we can estimate these coefficients via Monte-Carlo averages:
\BEA\label{Eqn:CYB}
%\begin{aligned}
\left[ \bm{C}_{{YZ}}\right] _{ks} \approx  \frac{1}{N}\sum_{i=1}^{N}\psi
_{k}\left( y_i\right) \varphi _{s}\left( \bm{z}_{i}\right),\quad\quad
\left[ \bm{C}_{{ZZ}}\right] _{sl} \approx  \frac{1}{N}%
\sum_{i=1}^{N}\varphi _{s}\left( \bm{z}_{i}\right) \varphi _{l}\left(
\bm{z}_{i}\right).
%\end{aligned}
\EEA
We should point out that if the weight $\hat{q}$ in $L^2(\mathcal{Z},\hat{q})$ is the sampling density of the data in $\mathcal{Z}$, since $\{\varphi_s\}$ is orthonormal under the corresponding inner product, then $\bm{C}_{ZZ}$ is an identity matrix. While a representation on this Hilbert space is desirable, finding the corresponding orthonormal basis for high-dimensional $\mathcal{Z}$ is computationally challenging. In addition to constructing the basis, the main computational cost arises when evaluating the estimated basis functions on new points $\hat{z}_t$ for future-time prediction as shown in the next section. To avoid these expensive computations, we will adopt simpler basis functions, namely the Hermite polynomial basis for low-dimensional $\mathcal{Z}$ and the proper orthogonal decomposition (POD) basis for high-dimensional $\mathcal{Z}$.

\subsection{Modeling the missing dynamics}

Given the pre-computed conditional density in \eqref{Eqn:tranker}, the closure modeling approach proposed in \eqref{abstractrom} requires estimating the following statistical quantities,
\begin{equation}\label{Exa}
\begin{aligned}
\hat{a}(\hat{x},\bm{\hat{z}}_t) &:= \mathbb{E}[a|\bm{\hat{z}}_t] := \int_\mathcal{Y} a(\hat{x},y) p(y|Z = \bm{\hat{z}}_t)\,dy, \\
\hat{B}(\hat{x},\bm{\hat{z}}_t) &:= \mathbb{E}[bb^\top |\bm{\hat{z}}_t] := \int_\mathcal{Y} b(\hat{x},y)b(\hat{x},y)^\top p(y|Z =\bm{\hat{z}}_t)\,dy.
\end{aligned}
\end{equation}
In the discussion below, we will just focus on the expectation of $a$ (the calculation of the expectation of $bb^\top$ will be similar). In our formulation, we set the weight $q$ in the Hilbert space $L^2(\mathcal{Y},q^{-1})$ to be the sampling density of the data in $\mathcal{Y}$. In particular, substituting \eqref{Eqn:tranker} into \eqref{Exa}, we obtain,
\BEA
\mathbb{E}[a|\bm{\hat{z}}_t] = \sum_{k,l=1}^\infty\int_{\mathcal{Y}}a(\hat{x},y)\psi _{k}\left( y\right)
q\left( y\right) dy\left[ \bm{C}_{{YZ}}\bm{C%
}_{{ZZ}}^{-1}\right] _{kl}\varphi _{l}\left( \bm{\hat{z}}_t\right)
= \sum_{l=1}^\infty A_{l}(\hat{x})\varphi _{l}\left( \bm{\hat{z}}_t\right),
  \label{Eqn:exp_gaussi}
\EEA%
where
\BEA
 A_{l}(\hat{x}) &:=& \sum_{k=1}^\infty \int_{\mathcal{Y}}a(\hat{x},y)\psi _{k}\left( y\right) q\left( y\right) dy \left[\bm{C}_{{YZ}}\bm{C}_{{ZZ}}^{-1}\right] _{kl} \nonumber \\
 &\approx &\frac{1}{N}\sum_{i=1}^N a(\hat{x},y_i) \sum_{k=1}^\infty \psi _{k}(y_i) \left[\bm{C}_{{YZ}}\bm{C}_{{ZZ}}^{-1}\right] _{kl} \nonumber \\
 &=&\frac{1}{N}\sum_{i=1}^N a(\hat{x},y_i)\sum_{k=1}^\infty \psi _{k}(y_i) \sum_{s=1}^\infty\left[\bm{C}_{{YZ}}\right]_{ks}\left[\bm{C}_{{ZZ}}^{-1}\right] _{sl} \nonumber \\
&\approx&\frac{1}{N^2}\sum_{i,j=1}^N a(\hat{x},y_i)\sum_{k=1}^\infty \psi _{k}(y_i) \sum_{s=1}^\infty\psi
_{k}\left( y_j\right) \varphi _{s}\left( \bm{z}_{j}\right) \left[\bm{C}_{{ZZ}}^{-1}\right] _{sl} \nonumber \\
&\approx&\frac{1}{N}\sum_{i=1}^N a(\hat{x},y_i) \sum_{s=1}^\infty \varphi _{s}\left( \bm{z}_{i}\right) \left[\bm{C}_{{ZZ}}^{-1}\right]_{sl} \label{coeffA}
\EEA
can be pre-computed. In this derivation, the second line is due to the Monte-Carlo average using data $y_i\sim q$, the fourth line above used \eqref{Eqn:CYB}, and the last line is due to the truncation in the summation of the index$-k$ up to order $N$, and
the fact that,
\BEA
\frac{1}{N}\sum_{k=1}^{N}\psi _{k}\left( y_{i}\right) \psi
_{k}\left( y_{j}\right) =\delta _{ij},\label{transpose}
\EEA
whenever $\{\psi_k\}$ is orthonormal in $L^2(\mathcal{Y},q)$, where the weight $q$ is exactly the sampling density of $\{y_i\}$. {\color{black} To see \eqref{transpose}, define an $N\times N$ matrix with components $A_{ij}=\psi_j(y_i)$, then the orthonormality condition means that $A^\top A=I$, where $I$ denotes an $N\times N$ identity matrix. Thus, \eqref{transpose} is the $(i,j)$th component of $AA^\top = I$.}
Since the resulting coefficients in \eqref{coeffA} are independent to $\psi_k(y)$, in practice, we only need to choose the basis $\varphi_l(\bm{z})$.

Notice that the resulting representation in \eqref{Eqn:exp_gaussi} arising from the proposed nonparametric formulation in \eqref{Eqn:tranker} is a parametric model {\color{black}when the summation term is truncated}, where the parametric ansatz is determined by how $\varphi_l$ depends on $\bm{z}$. For example, when $\mathcal{Z}$ is low-dimensional, we will consider Hermite polynomial basis functions for $\{\varphi _{l}\left( \bm{z}\right)\}_{l=1,\ldots, L}$ in a numerical example in Section~\ref{sec:ex1}. In this case, the resulting parametric model is a polynomial of degree$-L$ and the coefficients in $A_{l}(\hat{x})$ are directly estimated via the kernel embedding formula.

We should point out that when we use the Hermite polynomial basis, we set the weight $\hat{q}$ to be Gaussian with mean and covariance determined empirically from the training data $\{\bm{z}_i\}_{i=1,\ldots,N}$.
In our numerics, we also employ a regularization $\left( \bm{C}_{{ZZ}}+\lambda
\bm{I}\right) ^{-1}$ replacing  $\bm{C}_{{ZZ}}^{-1}$ in \eqref{coeffA}, with a small parameter $\lambda $ to compensate for the conditional density that is not in $\mathcal{H}$ (as suggested in \cite{Song2013IEEE,Song2009hilbert}). Basically, this regularization is the penalty of not building the appropriate RKHSs that respect the sampling distribution and geometry of the data.

For high-dimensional $\mathcal{Z}$, we will consider using the proper orthogonal decomposition (POD) as a basis for $%
\varphi _{l}\left( \bm{z}\right)$. Conceptually, this choice of basis corresponds to using an empirical covariance as the kernel in \eqref{intoperator} (see e.g., Chapter~5 of \cite{harlim2018} for more detailed discussion). Computationally, define a matrix $\bm{Z} \in\mathbb{R}^{N\times n_z}$, where the $i$th row consists of the training data, $\bm{z}_i -\bar{\bm{z}}\in \mathbb{R}^{n_z}$, centered about its empirical mean, $\bar{\bm{z}} = \frac{1}{N}\sum_{i=1}^N\bm{z}_i$, such that its row sum is zero. In this case, {\color{black}the function value $\varphi_j(\bm{z}_i)$ will be determined by the $(i,j)$th component of the
orthonormal matrix $\bm{U}$} defined as,
\begin{equation}
\bm{U}=\bm{ZV\Sigma }^{-1},  \label{Eqn:U}
\end{equation}%
where $\bm{Z}=\bm{U\Sigma V}^\top$ is the singular value
decomposition (SVD). These basis functions are called the Proper Orthogonal Decomposition (POD) modes or a discrete version of the Karhunen-Lo\`{e}ve basis expansion (see e.g., Chapter~5 of \cite{harlim2018}).

From the orthonormality of $\bm{U}$, we have $\bm{C}_{%
{ZZ}}=\bm{I}/N$ such that Eq. (\ref{Eqn:exp_gaussi}) can be further
simplified to,%
\begin{equation}
\mathbb{E}\left(a|\bm{\hat{z}}_t\right) =\sum_{i=1}^{N}\sum_{s=1}^{L}
a(\hat{x},y_i) \varphi _{s}\left( \bm{z}_{i}\right) \varphi _{s}\left(
\bm{\hat{z}}_t\right),  \label{Eqn:pcorht}
\end{equation}%
where we used $L$ basis functions. Suppose that $a(\hat{x},y)= y$, then Eq. (\ref{Eqn:pcorht}) can be equivalently rewritten in a matrix form as,
\begin{equation}
\mathbb{E}\left( Y|\bm{\hat{z}}_t\right) =\bm{Y}^{\top}%
\bm{U}\bm{U}_{\text{new}}^\top,  \label{Eqn:ker_matr}
\end{equation}%
{\color{black} where the matrix $\bm{Y}=[y_{1},\ldots
,y_{N}]^{\top}$ is $N\times n_y$ with $\left\{ y_i \right\} _{i=1}^{N}$ denoting the training data with dimension $n_y$. Here, the matrix $\bm{U}_{\text{new}}:=(\bm{\hat{z}}_t-\bm{\bar{z}})\bm{V\Sigma}^{-1}\in\mathbb{R}^{1\times L}$ is the Nystr\"{o}m extension for SVD \cite{nemtsov2016matrix}, whose components approximate the basis function values at a new point $\bm{\hat{z}}_t$, that is, $\bm{U}_{\text{new}}\approx\left[ \varphi _{1}\left(
\hat{\bm{z}}_t\right) ,\ldots ,\varphi _{L}\left( \hat{\bm{z}}_t\right) \right]$.} Substituting Eq. (\ref{Eqn:U}) into the
conditional expectation (\ref{Eqn:ker_matr}), we obtain
\begin{eqnarray}
\mathbb{E}\left( Y|\bm{\hat{z}}_t\right)  =\bm{Y}^\top%
\bm{{Z}}\bm{V}\bm{\Sigma }%
^{-1}\bm{\Sigma }^{-1\top}\bm{V}^\top(\bm{\hat{z}}_t - \bm{\bar{z}}%
)^\top
=\left( \bm{Y}^\top\bm{{Z}}\right) \left(
\bm{{Z}}^\top\bm{{Z}}\right) ^{-1}(\bm{\hat{z}}_t - \bm{\bar{z}}%
)^\top.  \label{Eqn:kern_reg}
\end{eqnarray}%
The formula in (\ref{Eqn:kern_reg}) is exactly a
linear regression between observations $\left\{ y_{i}\right\}
_{i=1}^{N}$ and $\left\{ \bm{z}_{i}\right\} _{i=1}^{N}$. This means that
the nonparametric RKHS representation reduces to the parametric linear
regression when POD bases are used to represent functions defined on the $\mathcal{Z}$ space. In the case where $\bm{z}_i := (\bm{x}_{i-m:i},\bm{y}_{i-n:i-1})\in\mathcal{Z}$, $n_z=m+n+1$, the resulting closure model in \eqref{Eqn:kern_reg} is nothing but a linear autoregressive model for variable $x$ with a linear non-autonomous variable $y$.

%In the next two sections, we will numerically demonstrate the predictive skill of this closure model to overcome systems without scale separation.

While the POD representation is convenient for high dimensional problems, we should point out these basis functions may not be adequate for systems with nonlinear and/or non-Gaussian nature. In fact, we will show in Section~\ref{sec:ex3} that the POD basis representation is not sufficient to recover the missing terms in a nonlinear system even when the invariant density is close to Gaussian. In this case, we will find that an additional {\color{black}Gaussian white} noise term can be used to compensate for the residual space (orthogonal to POD).

\section{\label{section3}A linear  Gaussian example}

In this section, we provide an intuitive argument for the choice of conditional density function $p(y_{t}|\bm{z}_t)$ in compensating the missing dynamical terms as proposed in \eqref{abstractrom}. Specifically, we will build our intuition for choosing variables $\bm{z}_t$ by studying the missing dynamics of an analytically tractable linear Gaussian problem
with and without temporal scale gaps. That is, we consider a linear multi-scale dynamical model,%
\begin{eqnarray}
dx &=&\left( a_{11}x+a_{12}y\right) dt+\sigma _{x}dW_{x},  \label{Eqn:eqn_x}
\\
dy &=&\frac{1}{\epsilon }\left( a_{21}x+a_{22}y\right) dt+\frac{\sigma _{y}}{%
\sqrt{\epsilon }}dW_{y},  \label{Eqn:eqn_y}
\end{eqnarray}%
for a slow variable $x\in \mathbb{R}$ and a fast variable $y\in \mathbb{R}$
\cite{gottwald2013role}. Here, $W_{x}$ and $W_{y}$ are independent Wiener
processes. The parameters $\sigma _{x},\sigma _{y}\neq 0$ and the
eigenvalues of the matrix
\begin{equation*}
A_\epsilon =\left(
\begin{array}{cc}
a_{11} & a_{12} \\
\frac{1}{\epsilon }a_{21} & \frac{1}{\epsilon }a_{22}%
\end{array}%
\right)
\end{equation*}%
are strictly negative, to assure the existence of a unique invariant joint
density $\peq \left( x,y\right) $. The parameter $\epsilon >0 $\ characterizes
the time-scale separation between variables $x$ and $y$. Moreover, we assume
the coefficient%
\begin{equation}
\widetilde{a}=a_{11}-a_{12}a_{22}^{-1}a_{21}<0, \quad a_{22}<0, \label{Eqn:atild_linear}
\end{equation}%
to assure that the leading-order slow dynamics supports an invariant measure $\hat{\rho}_{eq}(x)$.

When there is a time-scale gap, in the limit of $\epsilon\to 0$, the leading-order dynamics,
\begin{equation}
d\hat{x}_t=a \hat{x}_tdt+\sigma _{x}dW_{x},  \label{Eqn:linear_asympt}
\end{equation}%
with $a = \widetilde{a}$ as defined in \eqref{Eqn:atild_linear}, can be obtained by averaging the slow component of the vector field, $(a_{11}x+ a_{12}y)$, with respect to the invariant density $\rho_{\infty}(y;\hat{x}_t)$ of the fast dynamics in \eqref{Eqn:eqn_y} for a fixed $\hat{x}_t:=\hat{x}(t\tau)$. For this simple example, it is clear that $\rho_{\infty}(y;x) = \mathcal{N}(-a_{21}a_{22}^{-1}x,-.5\sigma_y^2a_{22}^{-1})$. The effective equation in \eqref{Eqn:linear_asympt}
is deduced using the averaging theory \cite{khasminskii:68,kurtz:75,papanicolaou1976some}, which approximates the density of the full dynamics as,
\BEA
p(x,y,t) = \hat{\rho}(x,t) \rho_{\infty}(y;x) + \mathcal{O}(\epsilon), \quad\quad t\geq 0,\label{averagingtheory}
\EEA
where $\hat\rho(x,t)$ denotes the evolution density corresponding to the leading-order dynamics.
First, we should point out that when the fast dynamics for $y$ in \eqref{Eqn:eqn_y} is not available, we have no information about the invariant density $\rho_{\infty}(y;x)$ and we also cannot generate samples of this density. Thus, $\tilde{a}$ is not computable since $a_{21}$ and $a_{22}$ are unknown.

Our proposed model in \eqref{abstractrom} for the closure is motivated by the following observation. Here, we first provide the theoretical validity of our closure model. Taking $t\to\infty$ in \eqref{averagingtheory},
the invariant density of the full dynamics can be approximated by that of the leading-order dynamics up to order-$\epsilon$, that is, $\peq(x,y) = \hat{\rho}_{eq}(x) \rho_{\infty}(y;x) + \mathcal{O}(\epsilon)$. Therefore,
\BEA
p(y|x) := \frac{\peq(x,y)}{\int_\mathcal{Y}\peq(x,y)\,dy } = \frac{\hat{\rho}_{eq}(x) \rho_{\infty}(y;x)+ \mathcal{O}(\epsilon) }{\hat{\rho}_{eq}(x) +\mathcal{O}(\epsilon)}  = \rho_\infty (y;x) + \mathcal{O}(\epsilon).\label{conditionalexpansion}
\EEA
This equation basically suggests that one can approximate $\rho_\infty (y;x)$ with the following conditional density $p(y|x)$.
For this linear Gaussian example, one can solve the Lyapunov equation of the full system in \eqref{Eqn:eqn_x}-\eqref{Eqn:eqn_y} for the equilibrium covariance matrix $S=(s_{ij})_{i,j=1,2}$ and deduce that $p(y|x) = \mathcal{N}(s_{21}s_{11}^{-1}x, s_{22}-s_{21}s_{11}^{-1}s_{12})$. Expanding the mean and variance statistics in terms of $\epsilon$, we obtain
\BEA
\mathbb{E}[Y|x] &:=& \int_{\mathcal{Y}} y p(y|x)\,dy = -a_{22}^{-1}a_{21}x + \mathcal{O}(\epsilon),\nonumber \\
\mathbb{E}[Y^2|x] &:=& \int_{\mathcal{Y}} y^2 p(y|x)\,dy = -\frac{\sigma_y^2}{2a_{22}} + \mathcal{O}(\epsilon),\nonumber
\EEA
which means that the order-$\epsilon$ expansion error in \eqref{conditionalexpansion} is in the sense of the mean and variance.

Averaging the slow Eq. (%
\ref{Eqn:eqn_x}) with respect to this conditional density, $p(y|\hat{x}_t) $, we obtain a  closure model of the form  \eqref{Eqn:linear_asympt} with
\begin{eqnarray}
a\hat{x}_t = \overline{a}\hat{x}_t =\int_{\mathcal{Y}} \left( a_{11}\hat{x}_t+a_{12}y\right) p\left(
y|\hat{x}_t\right) dy=\left( a_{11}+a_{12}s _{21}s_{11}^{-1}\right) \hat{x}_t = \tilde{a}\hat{x}_t + \mathcal{O}(\epsilon),\label{lina}
\end{eqnarray}
which means that the proposed closure obtained by averaging over $p(y_t|x_t)$ is consistent (up to an order-$\epsilon$ error) with the reduced model obtained from the classical averaging theory.
However, in general, such an analytical expression in \eqref{lina} will not be available since we have no access to $s_{21}$ and $s_{11}$. Numerically, we will approximate the conditional density, $p(y|x)$, by applying the kernel embedding of the conditional distributions discussed in the previous section on the training data set $\{x_i,y_i\}_{i=1}^N$. In this case, it is clear that $\bm{z}_t=x_t$ is the natural choice. In the remainder of this section, we will refer to this closure model as the ``RKHS $p(y_t|x_t)$".

Now we turn to the discussion of our closure model for large $\epsilon$. When there is no time-scale gap, i.e., $\epsilon=\mathcal{O}(1)$ is large, the approximation via the averaging theory is not valid, and thus, averaging over $p(y_t|x_t)$ will not work. In this case, let us consider $\bm{z}_t=\bm{{x}}_{t-m:t}$ such that our closure model is an average over a non-Markovian conditional density function $p(y_t|\bm{{x}}_{t-m:t})$. That is,
\BEA
d\hat{x}_t = (a_{11}\hat{x}_t+a_{12}\mathbb{E}\left[ Y|\bm{\hat{x}}_{t-m:t} \right])\,dt + \sigma_x dW_x(t),\label{modelwithmemory}
\EEA
where the conditional average is evaluated at a new data point $\bm{\hat{x}}_{t-m:t} := \Big(\hat{x}((t-m)\tau),\hat{x}((t-m+1)\tau),\ldots,\hat{x}(t\tau)\Big)$ for the time lag interval $\tau>0$, resulting from the integration of \eqref{modelwithmemory} at previous time steps. Since the random variables $Y$ of $y_t$ and $X$ of $\bm{x}_{t-m:t}$ are both Gaussian with mean zero and covariance,
\begin{equation}
\text{Cov}\left( \left[
\begin{array}{c}
{Y} \\
{X}%
\end{array}%
\right] ,\left[
\begin{array}{c}
{Y} \\
{X}%
\end{array}%
\right] \right)
:= \left[
\begin{array}{cc}
\Sigma _{11} & \Sigma _{12} \\
\Sigma _{21} & \Sigma _{22}%
\end{array}%
\right],  \label{Eqn:linear_SIG}
\end{equation}%
we can deduce that
\BEA
\mathbb{E}\left[ Y |\bm{\hat{x}}_{t-m:t}\right] =\Sigma _{12}\Sigma _{22}^{-1}\bm{\hat{x}}_{t-m:t}.  \label{Eqn:linear_Expt}
\EEA
When the covariance components $\Sigma_{12}$ and $\Sigma_{22}$ are empirically estimated from the training data, notice that \eqref{Eqn:linear_Expt} is identical to the conditional expectation with respect to the kernel embedding of the conditional distributions formulated using the POD basis in \eqref{Eqn:kern_reg}. More importantly, one can analytically show that the autocovariance function (ACV) of the proposed non-Markovian model in \eqref{modelwithmemory} with $m\to \infty$ agrees with the ACV of the $x-$component of the full model (see Appendix~\ref{app:ACF_linear} for the detailed proof of this statement). The consistency of the ACV prediction as well as the closure in \eqref{Eqn:linear_Expt} with the RKHS formulation in \eqref{Eqn:kern_reg} justifies the choice of $\bm{z}_t = \bm{{x}}_{t-m:t}$ when $\epsilon$ is large. In the numerics below, we will verify the robustness of the non-Markovian closure model resulted from this choice of $\bm{z}_t$ in terms of the short-time prediction skill and the long-time statistics of ACVs for any $\epsilon>0$.

\begin{figure}[tbp]
%\flushleft \hspace*{-7.2mm}
\centering \includegraphics[scale=0.5]{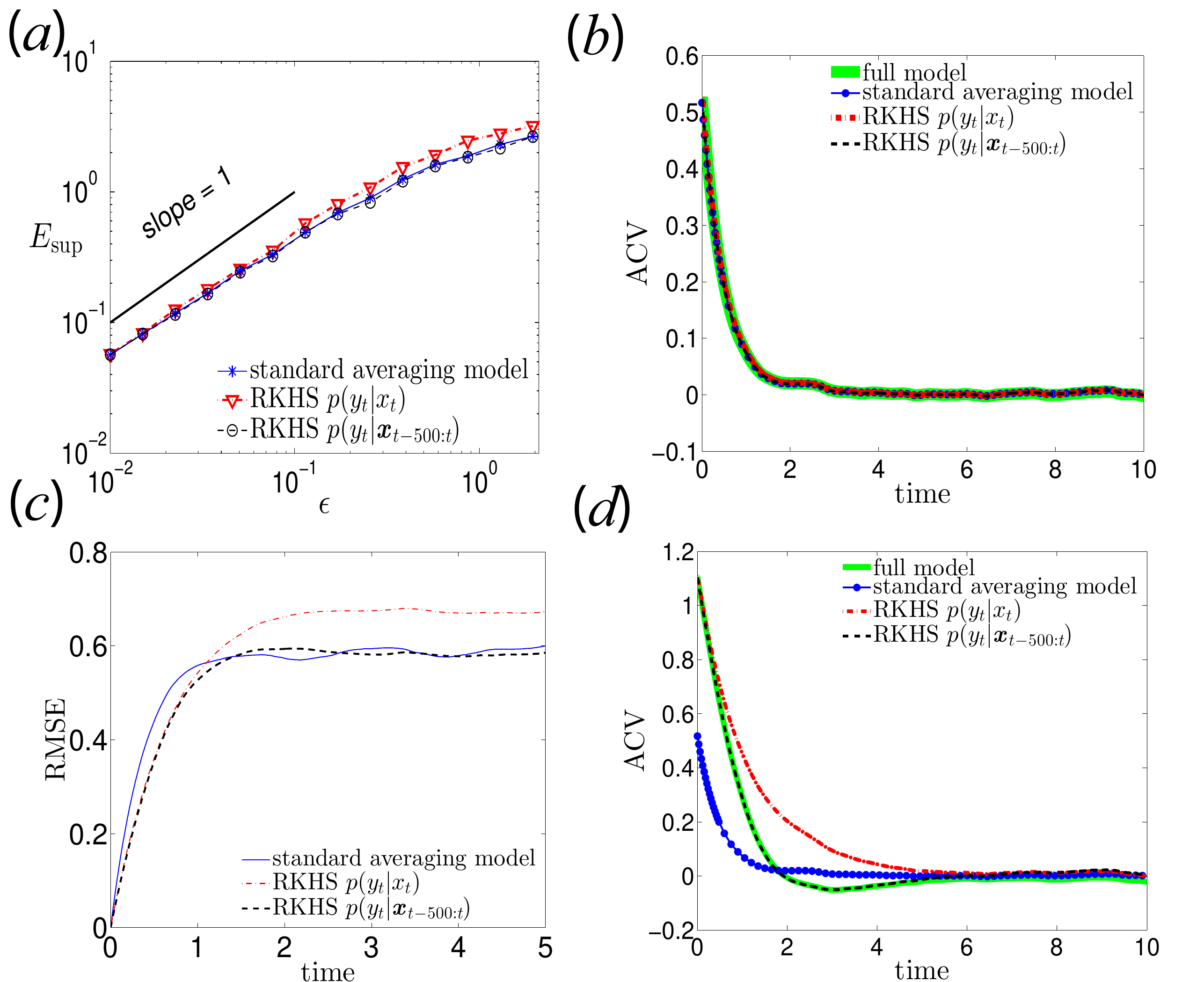}
%width=0.50\textwidth,height=0.30\textheight
\caption{(Color online) (a) Supremum errors $E_{\sup }$ as functions of
parameters $\protect\epsilon $, where $E_{\sup }\equiv \mathbb{E}\left(
\sup_{0\leq t\leq 40}\left\vert e\left( t\right) \right\vert ^{2}\right) $
with $e\left( t\right) =x(t) -\hat{x}(t) $. Here, $%
x\left(t\right)$ are solutions of the full model (\protect\ref{Eqn:eqn_x})-(%
\protect\ref{Eqn:eqn_y}) and $\hat{x}\left(t\right)$ are solutions of the closure
models. Trajectories are averaged over 100 realizations. The parameters are $%
a_{11}=a_{21}=a_{22}=-1$, $a_{12}=1$, and $\protect\sigma _{x}=\protect%
\sigma _{y}=\protect\sqrt{2}$. When $\protect\epsilon$ is small, the
solutions of all the closure models are pathwise convergent nearly on the
order of $\protect\epsilon$. (c) Comparison of RMSEs averaged over 1000
realizations for large $\protect\epsilon = 1.30$ regime. Comparison of ACVs for (b) the
small $\protect\epsilon = 0.01$ regime and (d) the large $\protect\epsilon =
1.30$ regime. }
\label{Fig5_LinearGaussian}
\end{figure}

In Figure~\ref{Fig5_LinearGaussian}, we compare our proposed closure model in \eqref{Eqn:linear_Expt} , which we will refer to as ``RKHS $p(y_t|\bm{x}_{t-m:t})$'', with the standard averaging model in \eqref{Eqn:linear_asympt} with $a=\widetilde{a}$ and the RKHS $p(y_t|x_t)$ as well. In this numerical simulation, we build the closure models using the simulated data at discrete time step $\tau=0.01$. When $\epsilon $ is small, one can observe the pathwise convergence of the solutions
of the closure models to those of the full model (\ref{Eqn:eqn_x})-(\ref{Eqn:eqn_y})
 [Fig. \ref{Fig5_LinearGaussian}(a)]. For small $\epsilon =0.01$,
the ACVs of all the closure models are in good agreement with the ACV of the
full model [Fig. \ref{Fig5_LinearGaussian}(b)]. These results agree with the invariant manifold theory for small $\epsilon $ \cite%
{pavliotis2008multiscale}. However, when $\epsilon $ is large, the short-time predictions and the long-time ACVs become
quite different among the three closure models [Figs. \ref{Fig5_LinearGaussian}(c) and (d)]. In term of short-time predictions,
the closure model (\ref{modelwithmemory}) with $m=500$ memory terms provides
a slightly better RMSE than the other two closure models [Fig. \ref{Fig5_LinearGaussian}(c)]. In term of long-time
statistics, only the closure model (\ref{modelwithmemory}) with long memory
terms produces an accurate approximation of the ACV, whereas the other
two closure models do not [Fig. \ref{Fig5_LinearGaussian}(d)]. This consistency of ACVs can be verified explicitly as we mentioned before (see Appendix~\ref{app:ACF_linear}).

The analysis over this simple example shows that the proposed modeling framework using the kernel embedding of the conditional density formulation provides accurate short-time predictions and consistent long-term statistical recoveries in the limit of the memory length $m\to\infty$. This consistency is robust whether the underlying full system has or does not have any temporal scale gap. Using this result as a guideline, a natural extension for compensating missing components in nonlinear systems is to consider $\bm{z}_t := (\bm{x}_{t-m:t},\bm{y}_{t-n:t-1})$, that allows for the missing dynamical components to also depend on the history of $y$ in addition to that of $x$. In practice, the key parameters which will be determined case-by-case are the memory length, $m$ and $n$, as we shall see in the nonlinear examples in the next section.

\section{\label{sec:examples} Nonlinear examples}

In this section, we study the short-time prediction and long-time
statistical properties of two nonlinear examples: the Lorenz-96
(L96) model \cite{Lorenz1996} possessing a short memory effect and the truncated Burgers-Hopf (TBH) model \cite%
{majda2002statistical,majda2005information,majda2006stochastic,majda2000remarkable}
possessing a long memory effect.

\subsection{\label{sec:ex1} Two-layer Lorenz-96 model}

Consider the two-layer Lorenz-96 (L96) model \cite{Lorenz1996},
\begin{equation}\label{Eqn:Lorenz}
\begin{aligned}
\dot{X}^{k} &=X^{k-1}\left( X^{k+1}-X^{k-2}\right) -X^{k}+F+B^{k}, \\
\dot{Y}^{j,k} &=\frac{1}{\varepsilon }\left[ Y^{j+1,k}\left(
Y^{j-1,k}-Y^{j+2,k}\right) -Y^{j,k}+h_{y}X^{k}\right] ,
\end{aligned}
\end{equation}%
for $k=1,\ldots ,K,$ and $j=1,\ldots ,J$, where each relevant variable $%
X^{k} $ is coupled to $J$ irrelevant variables $Y^{j,k}$, and
\begin{equation}
B^{k}=\frac{h_{x}}{J}\sum_{j=1}^{J}Y^{j,k}.  \label{Eqn:Lorenz_Bk}
\end{equation}%
The indices of the variables $X^{k}$ and $Y^{j,k}$ are cyclic, $X^{k} =X^{k+K}, Y^{j,k} =Y^{j,k+K}, Y^{j+J,k} =Y^{j,k+1}$.
%The formulation of the model here is equivalent to the original one in Ref. \cite{Lorenz1996} (see, e.g., \cite%{fatkullin2004computational,crommelin2008subgrid}).
The parameters are taken
to be $K=18$, $J=20$, $F=10$, $h_{x}=-1$, and $h_{y}=1$ \cite%
{crommelin2008subgrid}. The parameter $\varepsilon $ characterizes the time
scale separation between the relevant component $X^{k}$ and the irrelevant component $Y^{j,k}$. In this example,
we will show the results for a small $\varepsilon =1/128$
and a large $\varepsilon =0.5$ (the large $%
\varepsilon =0.5$ regime was studied in \cite%
{crommelin2008subgrid,fatkullin2004computational,kwasniok2012data,lu2017accounting}%
). We integrate the full L96 model using a 4th-order Runge-Kutta method for $%
10^{3}$ time units with a time step $\delta t=0.001$. We observe the
trajectories of the variables $(X^{k},B^{k})$ every 10 time steps, so that
the observation time step is $\tau =0.01$ and the dataset contains $N=10^{5}$
observation points.

In the following numerical simulations, we compare our proposed closure RKHS models with the deterministic parametric formulation suggested by Wilk's method \cite{wilks2005effects}. In particular, the Wilk's deterministic parameterization scheme is a closure model obtained by fitting the data $\{(X^{k}_i,B^{k}_i)\}_{i=1}^N$ with the following polynomial,
\begin{equation}
B^{k}=b_{0}+b_{1}X^{k}+b_{2}\left( X^{k}\right) ^{2}+b_{3}\left(
X^{k}\right) ^{3}+b_{4}\left( X^{k}\right) ^{4}+b_{5}\left( X^{k}\right)
^{5}.  \label{Eqn:Wilks_5}
\end{equation}
We should point out that if we are restricted to only observing $\{X^k_i\}$, then $\{B^{k}_i\}$ are the identifiable components that can be extracted,  for example, using a likelihood maximum estimate \cite{kkg:05,lu2017data,wilks2005effects} or an adaptive Bayesian filtering \cite{bh:16jcp}, as we pointed out in the introduction. The key point is that we cannot extract the detailed components $Y^{j,k}$ if the fast dynamical components in \eqref{Eqn:Lorenz} are unknown and, in fact, we are not interested in constructing a closure model by averaging over conditional density that depends directly on $Y^{j,k}$ since this can be very expensive. Instead, we will consider a closure model based on averaging over the conditional density $p\left( B_{t}^{k}|X_{t}^{k}\right) $ for small $\epsilon$, where $B_{t}^{k}:=B^{k}\left( t\tau \right)$ and $X_{t}^{k}:=X^{k}\left( t\tau \right) $.  For the large $\varepsilon $ regime, we will consider $p\left(B_{t}^{k}|X_{t}^{k},B_{t-1}^{k}\right)$. While conditioning to other variables (e.g., spatial neighbors of $X^k$ or $B^k$ or longer temporal history) can be considered, we do not find any meaningful improvement over the results that are presented below. These densities will be constructed using the kernel embedding formulation discussed in Section~\ref{section2} for each $k$; connecting to the notation in the previous section, $y_t :=B^k_t$ and $\bm{z}_t$ is either $X_{t}^{k}$ or $(X_{t}^{k},B_{t-1}^{k})$.
{\color{black} To clarify, the full problem in \eqref{Eqn:Wilks_5} is $K+KJ = 18+18\times 20=378$ dimension, and the closure model for the missing components, $\{Y^{j,k}\}_{k=1,\ldots,K, j=1,\ldots, J}$, is defined through a set of either one-dimensional or two-dimensional conditional densities; i.e., for each $k$, the density takes either $X_{t}^{k}$ or $(X_{t}^{k},B_{t-1}^{k})$ as inputs. Since these densities are low-dimensional functions with respect to the conditional variables (either $X_{t}^{k}$ or $(X_{t}^{k},B_{t-1}^{k})$),} we will represent the kernel embedding formula in \eqref{Eqn:tranker} using the Hermite polynomials, {\color{black} expansion truncated at order $L=50$ for each memory term}.

\begin{figure}[tbp]
%\flushleft \hspace*{-7.2mm}
\centering \includegraphics[scale=0.5]{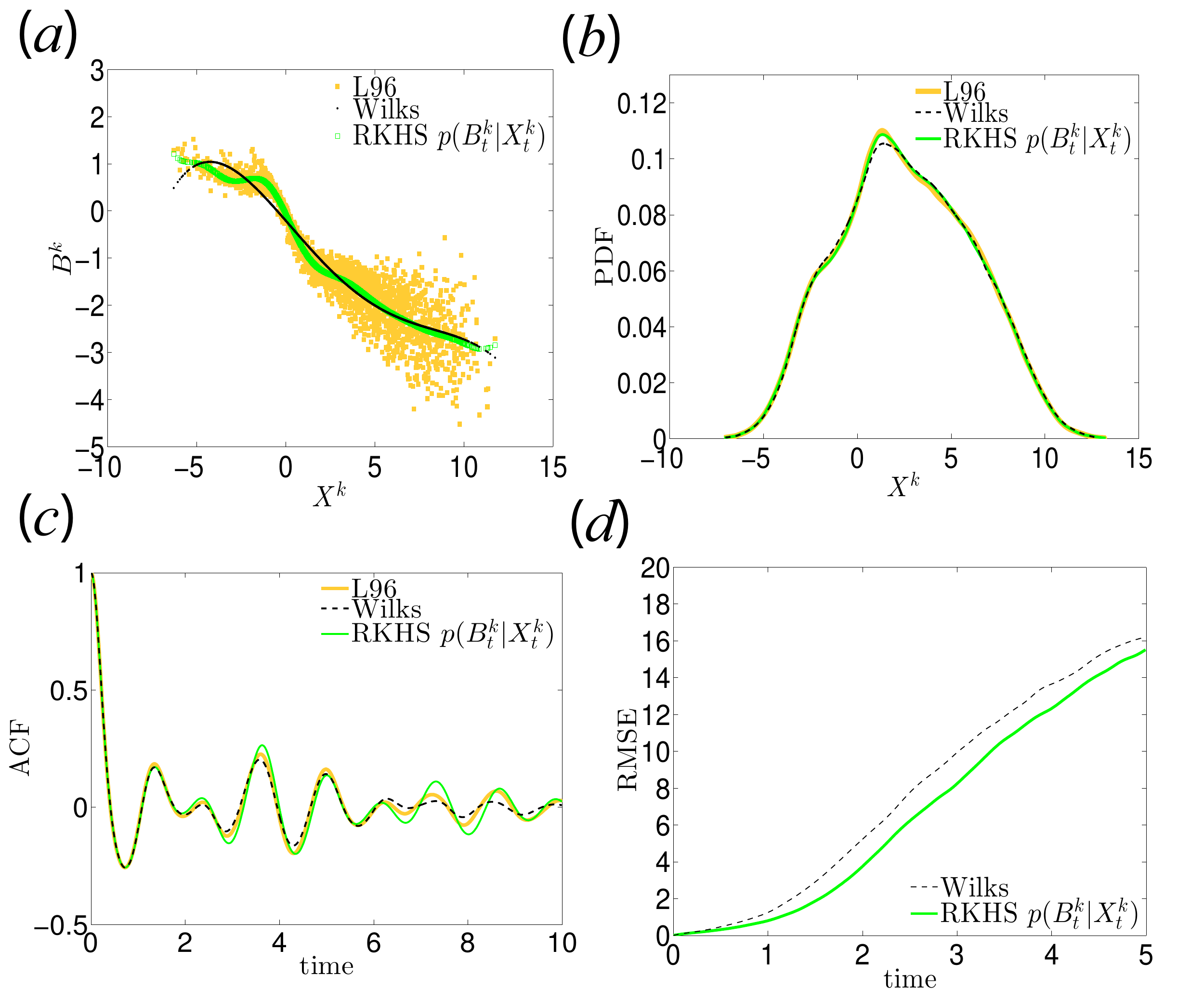}
%width=0.50\textwidth,height=0.30\textheight
\caption{(Color online) Long-time statistics and short-time predictions for
the small $\protect\epsilon=1/128$ regime of the L96 model. (a) The yellow
dots are the scatter plot of $B^{k}$ vs. $X^{k}$ for the full L96
model. The black dots are the fifth-order polynomial fit used for the
deterministic parametrization of $B_{k}$ using Wilks's method \protect\cite%
{wilks2005effects}. The green squares are the closure model using the conditional density
$p\left( B^{k}_t |X^{k}_t \right) $.
Comparison of (b) PDFs and (c) ACFs among the full L96 model and the
closure models. (d)
Comparison of RMSEs from ensemble averages. The number of ensembles is 1000
where each ensemble corresponds to an initial state.}
\label{Fig1_smalleps}
\end{figure}

To validate the proposed approach, we compare  the short-time predictions and long-time statistics of the $X^k$-components between the full model and the closure models. Particularly, we compare several standard long-time statistical quantities as in \cite{crommelin2008subgrid,lu2017data}:
\begin{itemize}
\item The probability density function (PDF) for $X^{k}$.
\item The autocorrelation function (ACF) for $X^{k}$, $%
\left\langle X_{t}^{k}X_{0}^{k}\right\rangle /\left\langle
X_{0}^{k}X_{0}^{k}\right\rangle $, where $\langle \cdot \rangle$ denotes the temporal average over $N=10^5$ data points.
\item The cross-correlation function (CCF) between $X^{k}$ and $%
X^{k+1}$, $\left\langle X_{t}^{k}X_{0}^{k+1}\right\rangle /\left\langle
X_{0}^{k}X_{0}^{k}\right\rangle $.
\item The mean wave amplitude $\left\langle \left\vert
u^{m}\right\vert \right\rangle $, for $m=0,\ldots ,K/2$, where $u^{m}$ is
the Fourier transform of $X^{k}$.
\item The wave variance $\left\langle \left\vert
u^{m}-\left\langle u^{m}\right\rangle \right\vert ^{2}\right\rangle $.
\end{itemize}

\noindent For the PDFs, ACFs, and CCFs, we plot the average over all $%
k=1,\ldots ,K$. For small $\varepsilon =1/128$, we only show the results for
the PDFs and ACFs. To assess the short-time prediction skill, we calculate the
root-mean-square error (RMSE) and the anomaly correlation (ANCR), where the
RMSE measures the difference between the true trajectory and the forecast
trajectory whereas the ANCR measures the correlation between them \cite%
{crommelin2008subgrid}. The definitions of RMSE and ANCR are the same as
those in \cite{crommelin2008subgrid}. We take the average using the
data from $1000$ different ensembles, each starting from a different initial
state over five time units.

We first report the small $\varepsilon  =1/128 $\ regime of the
L96 model. Figure \ref{Fig1_smalleps}(a) displays the scatter plot of $%
B_{t}^{k}$ vs. $X_{t}^{k}$ for the full L96 model, the polynomial fit (\ref%
{Eqn:Wilks_5})\ for the deterministic parametrization of $B_{k}$ (Wilks's
method), and the expectation $\mathbb{E}\left[ B_{t}^{k}|X_{t}^{k}\right] $
using the RKHS representation (method referred to as the RKHS $p\left( B_{t}^{k}|X_{t}^{k}\right) $). For long-time statistics, one can see from
Figs. \ref{Fig1_smalleps}(b) and \ref{Fig1_smalleps}(c) that the PDFs and
ACFs for $X^{k}$ can be well reproduced by both closure models. For
short-time predictions, one can see from Fig. \ref{Fig1_smalleps}(d) that
the RKHS $p\left( B_{t}^{k}|X_{t}^{k}\right) $ provides a  better approximation of the trajectory
compare to the Wilks's deterministic parametrization scheme. These results can be
expected due to the validity of the classical averaging theory on dynamical
systems with time-scale separation (small $\varepsilon $\ regime) \cite%
{pavliotis2008multiscale}.

\begin{figure}[tbp]
%\flushleft \hspace*{-7.2mm}
\centering \includegraphics[scale=0.5]{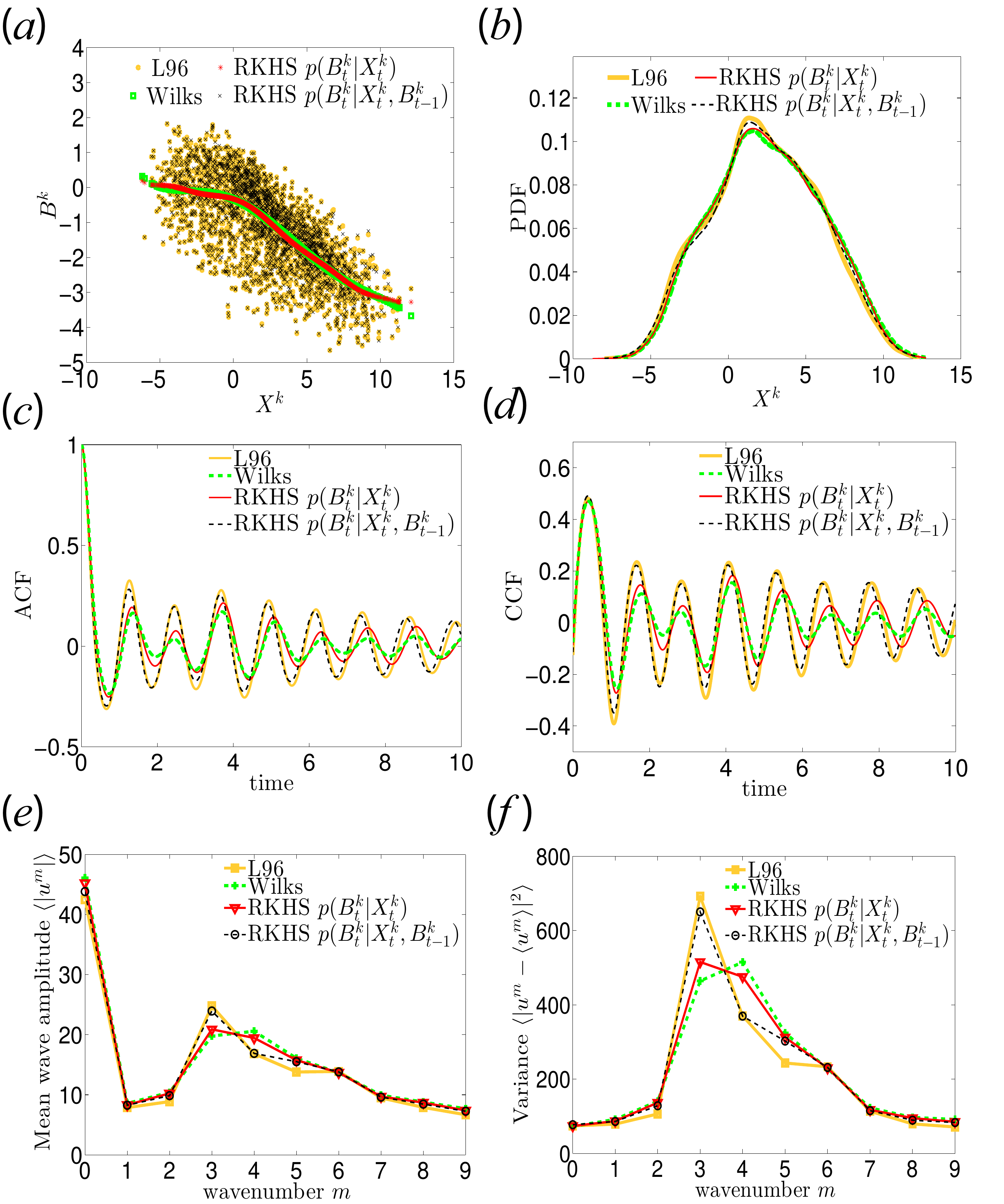}
%width=0.50\textwidth,height=0.30\textheight
\caption{(Color online) Long-time statistics and short-time predictions for
the large $\protect\epsilon=0.5$ regime of the L96 model. (a) The yellow dots
are the scatter plot of $B^{k}$ vs. $X^{k}$ for the full L96 model.
The green squares are the fifth-order polynomial fit using Wilks's method
\protect\cite{wilks2005effects}. The red asterisks and black crosses correspond
to the closure models using the conditional densitys $p\left( B^{k}_t |X^{k}_t \right) $
and $p\left( B^{k}_t |X^{k}_t
,B^{k}_{t-1} \right) $, respectively.  Comparison of (b) PDFs, (c) ACFs, (d) CCFs, (e) mean wave
amplitudes, and (f) wave variances among the full model and closure models.}
\label{Fig2_RSME_smalleps}
\end{figure}

We now report the L96 model for the large $\varepsilon  =0.5 $
regime in which there is no significant time-scale separation between the
relevant,  $X^{k}$, and irrelevant variables, $B^{k}$. By comparing
Fig. \ref{Fig1_smalleps}(a) and \ref{Fig2_RSME_smalleps}(a), one can see
that the patterns of the scatter plots differ substantially between the
small and large $\varepsilon $ regimes. Specifically, the scatter plot for
the large $\varepsilon $ regime is much broader in $B^k$ direction compare to that for the small $%
\varepsilon $ regime. This indicates that when $\varepsilon $ is small, the
irrelevant (fast) variable significantly relies on the relevant (slow)
variable. When $\varepsilon $ becomes large, such dependence of irrelevant
variable $B^{k}$\ on the relevant variable $X^{k}$ reduces.

\comment{
For the large $%
\varepsilon $ regime, the basic idea is to increase such dependence by
adding more memory terms. In particular, we add one more memory term $%
B_{t-1}^{k}$, that is, we use the RKHS representation for the transition
kernel $p\left( B_{t}^{k}|X_{t}^{k},B_{t-1}^{k}\right) $.}

\begin{figure}[tbp]
%\flushleft \hspace*{-7.2mm}
\centering \includegraphics[scale=0.5]{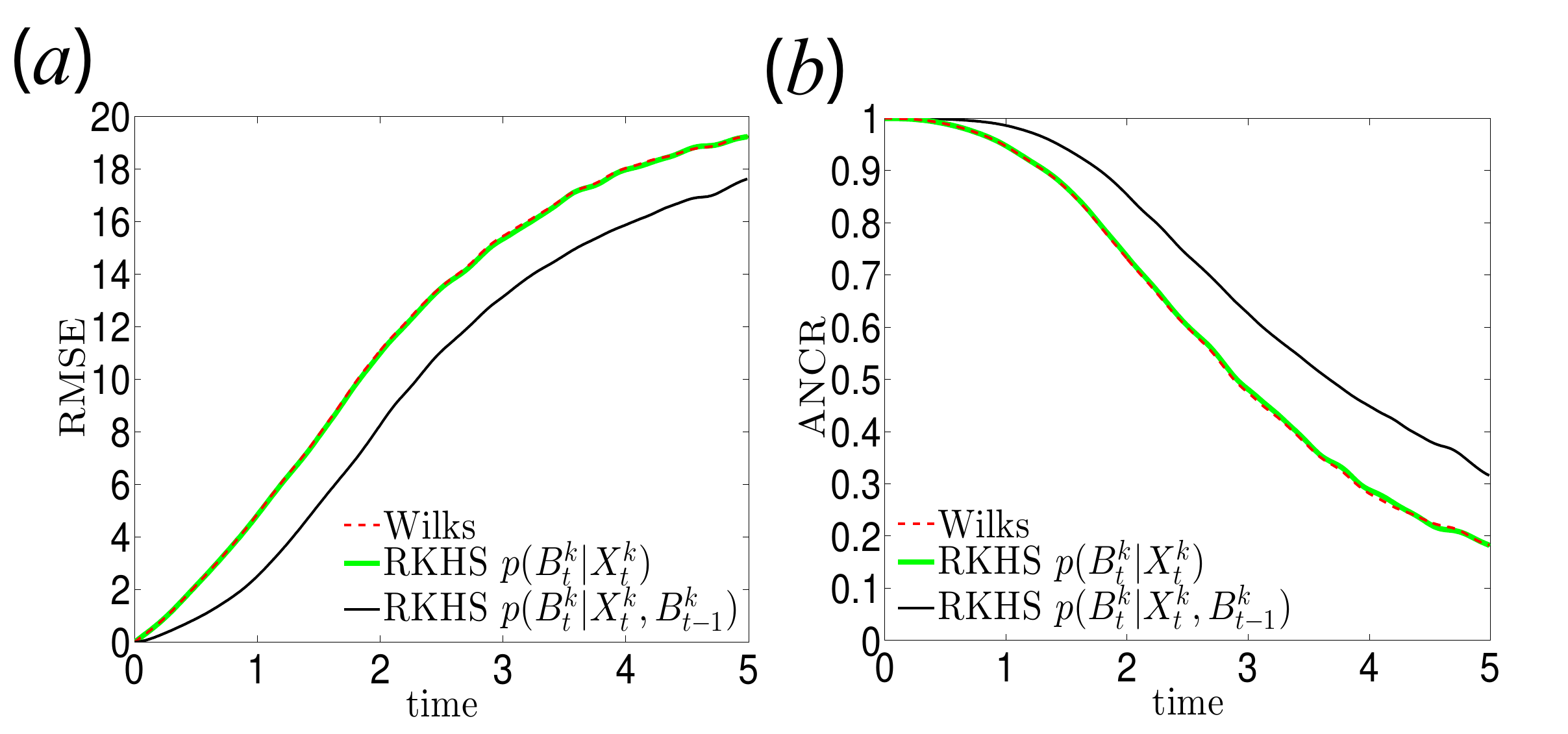}
%width=0.50\textwidth,height=0.30\textheight
\caption{(Color online) Comparison of (a) RMSEs and (b) ANCRs for the large $%
\protect\epsilon=0.5$ regime. The number of ensembles is 1000 where each
ensemble corresponds to an initial state. }
\label{Fig3_RMSE_largeeps}
\end{figure}

For large $\varepsilon =0.5$, one can observe from Fig.~\ref{Fig2_RSME_smalleps}%
(a) that the RKHS representation of $\mathbb{E}\left[
B_{t}^{k}|X_{t}^{k},B_{t-1}^{k}\right] $ can nearly reproduce the scatter
plot of the full model, whereas the Wilks's deterministic parametrization
scheme and the RKHS representation $\mathbb{E}\left[ B_{t}^{k}|X_{t}^{k}%
\right] $\ cannot. The PDFs for $X^{k}$ of the full model can be reproduced
by all the closure models [Fig. \ref{Fig2_RSME_smalleps}(b)]. For the other
long-time statistics, ACFs, CCFs, mean wave amplitudes, and wave variances
can be well reproduced only by the closure model using the conditional density $%
p\left( B_{t}^{k}|X_{t}^{k},B_{t-1}^{k}\right) $ [Figs. \ref{Fig2_RSME_smalleps}%
(c)(d)(e)(f)]. Notice also the significant improvement in terms of short-time predictions using the RKHS $p\left( B_{t}^{k}|X_{t}^{k},B_{t-1}^k\right) $ (smaller RMSE and higher ANCR) over the Wilks's method and the RKHS $p\left( B_{t}^{k}|X_{t}^{k}\right) $ as shown in Fig.~\ref{Fig3_RMSE_largeeps}.
%However, these statistics cannot be well approximated by the reduced models using the deterministic parametrization scheme or using the transition kernel $p\left( B_{t}^{k}|X_{t}^{k}\right) $\ [Figs. \ref% {Fig1_smalleps}(c)(d)(e)(f)]. These results suggest that when $\varepsilon $ becomes large, it would be better to add more memory terms in the transition kernel.

\begin{figure}[tbp]
%\flushleft \hspace*{-7.2mm}
\centering \includegraphics[scale=0.5]{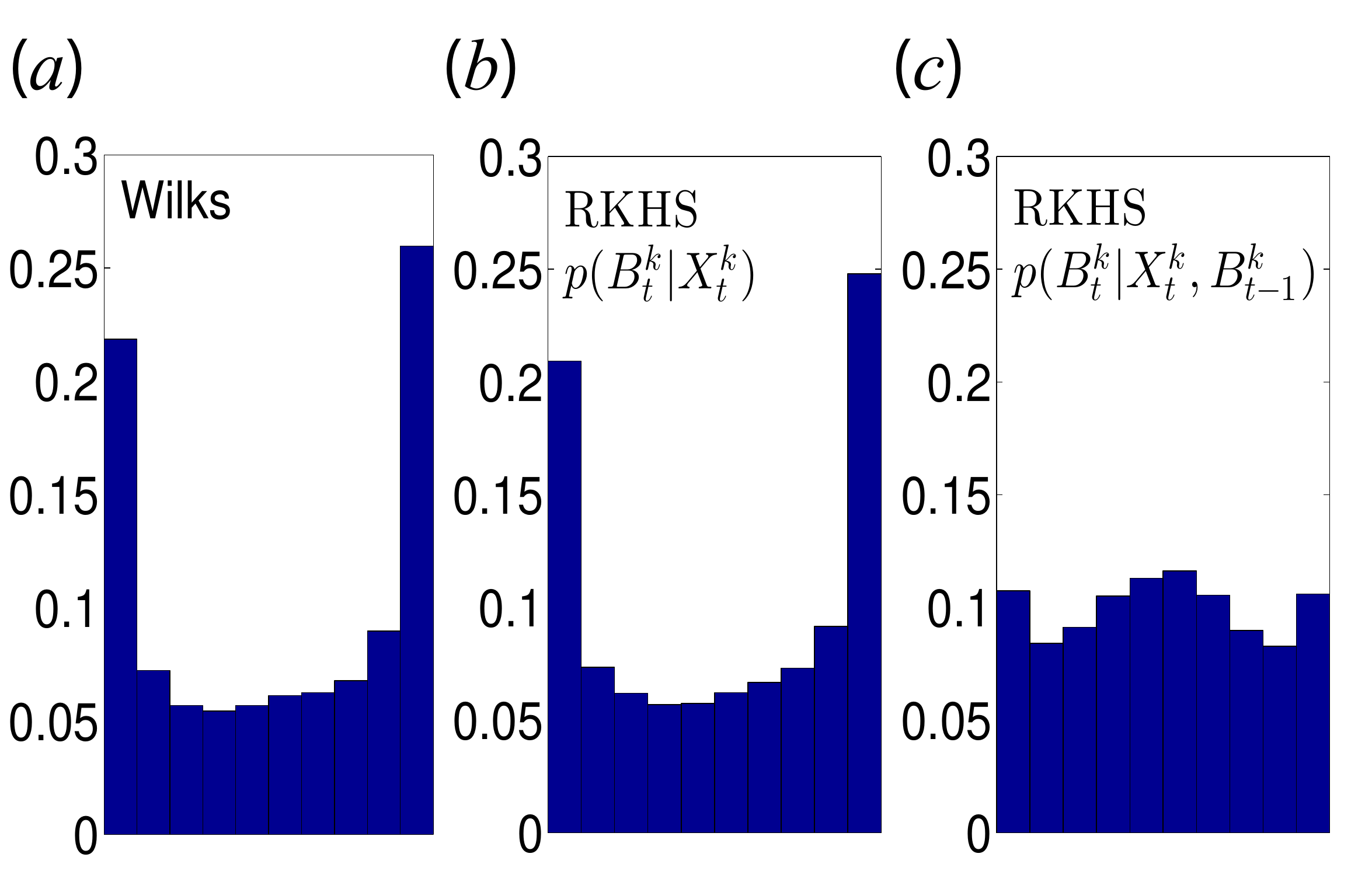}
%width=0.50\textwidth,height=0.30\textheight
\caption{(Color online) Rank histograms for closure models with ensemble
members $N_{\mathrm{ens}}=9$ at lead time $T=2$. Ideally, the
rank histogram is nearly flat. The rank histogram of the closure model using
$p\left( B^{k}_t |X^{k}_t
,B^{k}_{t-1} \right) $
is close to be flat, whereas rank histograms of Wilks's method and the closure model using
$p\left( B^{k}_t |X^{k}_t \right) $  exhibit U-shape distributions. }
\label{Fig4_Histrank}
\end{figure}

To determine the reliability of the ensemble forecasts, we also calculate the rank histograms from
an ensemble of integrations \cite{hamill2001interpretation}. A rank histogram is obtained by repeatedly tallying
the rank of the true observation relative to the sorted $N_{\mathrm{ens}}$%
-member ensemble \cite{hamill2001interpretation}. We use the same method as
in \cite{crommelin2008subgrid}. For every initial state $X_{t_{0}}^{k}$, we
do $N_{\mathrm{ens}}$ integrations of the closure models over the lead $T$
time units starting from the $X_{t_{0}}^{k}$ plus a small random
perturbation. The random perturbations are Gaussian distribution with mean
zeros and standard deviation $0.15$. We sort the $N_{\mathrm{ens}}+1$ values
for $X_{t}^{k}$\ for each grid point $k$ and time $t$ from the ensemble
members and the full L96 model. Figure \ref{Fig4_Histrank} displays the rank
histograms for all the closure models with $N_{\mathrm{ens}}=9$ at lead time
$T=2$. An ideal rank histogram is flat. One can see that the rank histogram
by the RKHS $p\left(
B_{t}^{k}|X_{t}^{k},B_{t-1}^{k}\right) $ is close to be flat, whereas rank
histograms by Wilks's deterministic parametrization scheme and the RKHS
$p\left( B_{t}^{k}|X_{t}^{k}\right) $
exhibit U-shape distributions. Therefore, the closure model with $p\left(
B_{t}^{k}|X_{t}^{k},B_{t-1}^{k}\right) $ performs better than the other two
closure models.

\subsection{\label{sec:ex3}The truncated Burgers-Hopf (TBH) model}

Consider the truncated Burgers-Hopf (TBH) model \cite%
{majda2002statistical,majda2005information,majda2006stochastic,majda2000remarkable}%
, which is described by a system of quadratic nonlinear equations for the complex
Fourier modes, $u^{k}$, with $u^{-k}=(u^{k})^{\ast }$ for $1\leq \left\vert
k\right\vert \leq \Lambda $,
\begin{equation}
\frac{du^k}{dt}=-\frac{\mathrm{i}k}{2}\sum_{\substack{ k+p+q=0  \\ 1\leq
\left\vert p\right\vert ,\left\vert q\right\vert \leq \Lambda }}(u^{p})^{\ast
}(u^{q})^{\ast }.  \label{Eqn:TBH_eq}
\end{equation}%
This model is a Galerkin truncation of the inviscid Burgers equation on Fourier modes and we should point out that the dynamics of the truncated system is totally different from the inviscid Burgers equation. Particularly, the TBH exhibits intrinsic stochastic dynamics with ergodic behavior in a large deterministic system \cite%
{majda2002statistical,majda2005information,majda2006stochastic,majda2000remarkable}. We are interested in estimating the TBH model's first Fourier mode given only the dynamical component of this mode,
\begin{equation}
\frac{du^{1}}{dt}=-\mathrm{i}(u^{1})^{\ast }u^{2}+F,  \label{Eqn:TBH_reduced}
\end{equation}
where $u_2$ denotes the second Fourier mode and $F$ denotes the forcing component
obtained by subtracting $-\mathrm{i}(u^{1})^{\ast }u^{2}$ from the right hand
side of Eq. (\ref{Eqn:TBH_eq}), that is,
\begin{equation}
F=-\frac{\mathrm{i}}{2}\sum_{\substack{ 1+p+q=0 \\ 2\leq \left\vert
p\right\vert ,\left\vert q\right\vert \leq \Lambda }}(u^{p})^{\ast
}(u^{q})^{\ast }.  \label{Eqn:TBH_force}
\end{equation}%
While $u^2$ and $F$ may be identifiable from observing $u^1$ alone, in our experiment below, we assume that we are given the data set of $\{u^1_{i},u^2_{i},F_i\}_{i=1}^N$. We should point out that this model has an equipartition energy, that is, all of the Fourier modes in TBH have the same variances, and the first Fourier mode (which is of our interest) possesses the longest autocorrelation time and the largest statistical memory \cite{majda2012physics}, which makes this example a tough test problem.

To compensate for the missing dynamics in (\ref{Eqn:TBH_reduced}), we substitute the
irrelevant variables $u^{2}$ and $F$ with their conditional expectations. In this case,
the closure model involves $p(y|\bm{x}_{t-m:t-1},\bm{y}_{t-n:t-1})$, where the
irrelevant variable $y$ is one of  $\{ u^{2,{\rm Re}}, u^{2,{\rm Im}}, F^{{\rm Re}}, F^{{\rm Im}}\}$,
and the relevant variable $\bm{x}$ is one of $\{u^{1,{\rm Re}},u^{1,{\rm Im}}\} $ such that $\bm{y}$ and $\bm{x}$ are
both real or both imaginary parts. In particular, we employ the RKHS formulation to construct four conditional densities: $p(u^{2,{\rm Re}}_t| \bm{u}^{1,{\rm Re}}_{t-m:t-1}, \bm{u}^{2,{\rm Re}}_{t-n:t-1})$, $p(u^{2,{\rm Im}}_t| \bm{u}^{1,{\rm Im}}_{t-m:t-1}, \bm{u}^{2,{\rm Im}}_{t-n:t-1})$, {\color{black} $p(F^{{\rm Re}}_t| \bm{u}^{1,{\rm Re}}_{t-m:t-1}, \bm{F}^{{\rm Re}}_{t-n:t-1})$, and $p(F^{{\rm Im}}_t| \bm{u}^{1,{\rm Im}}_{t-m:t-1}, \bm{F}^{{\rm Im}}_{t-n:t-1})$.} For the forcing $F$, an additional
Gaussian noise term is added to compensate for the residual space. Since the conditional states are high-dimensional (when $m,n$ are large), the conditional
expectations over these densities are represented using the POD bases as in (\ref{Eqn:kern_reg}).

\begin{figure}[tbp]
%\flushleft \hspace*{-7.2mm}
\centering \includegraphics[scale=0.4]{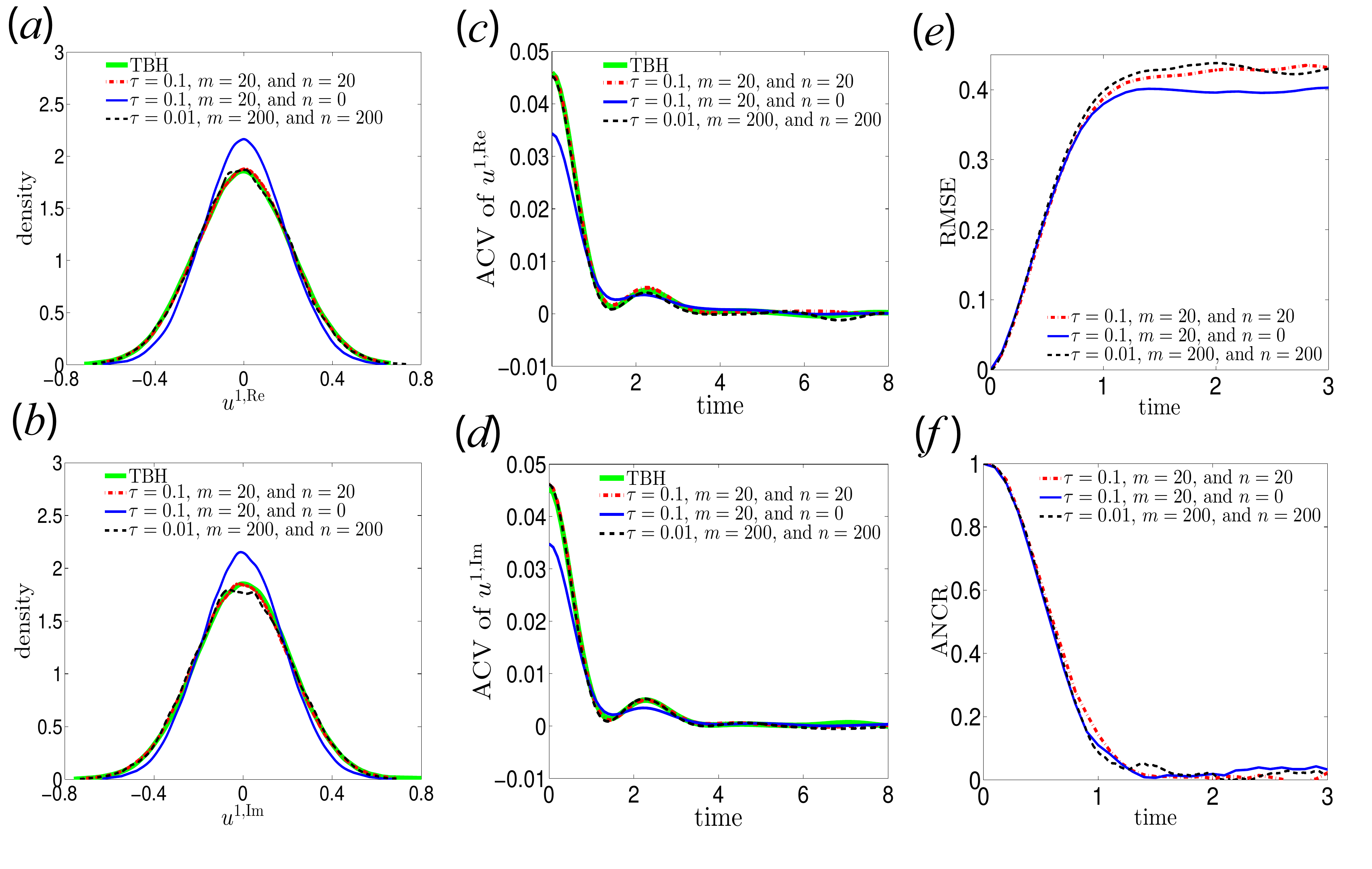}
%width=0.50\textwidth,height=0.30\textheight
\caption{(Color online) Long-time statistics and short-time predictions for
the TBH model.
Comparison of PDFs of (a) $ u^{1,{\rm Re}}$ and (b) $
 u^{1,{\rm Im}}$. Comparison of ACVs of (c) $ u^{1,{\rm Re}}$ and (d) $
 u^{1,{\rm Im}}$. Comparison of (e) RMSEs and (f)
ANCRs. The closure models use the conditional density $p(y|\bm{x}_{t-m:t-1},\bm{y}_{t-n:t-1})$ with
different observation
time step $\tau$ and number of memory terms $m$ and $n$ [see text].}
\label{Fig6_TBH}
\end{figure}

To conduct this numerical experiment, the training dataset is generated from the full TBH model (\ref{Eqn:TBH_eq}%
), where $F$ is calculated by Eq. (\ref{Eqn:TBH_force}). We integrate the full TBH model for $10^{4}$ time units with time step $\Delta t=10^{-3}$. We store the data at every $0.01$ time unit and thereafter the dataset contains $10^{6}$ points for all $u^{k}$. We compare the results generated by the full TBH model (\ref{Eqn:TBH_eq})\
and the closure models, resulted by averaging the partial dynamics in (\ref{Eqn:TBH_reduced}) over the pre-trained conditional densities. In this example, we
consider the full TBH model (\ref{Eqn:TBH_eq}) in a high-energy regime with $%
\beta =10$ and $\Lambda =50$ as in \cite{majda2000remarkable}.
Here, $\Lambda $ denotes number of modes in Eq. (\ref{Eqn:TBH_eq}) and $\beta
=\Lambda /\overline{E}$ with $\overline{E}$ being the mean energy per mode. {\color{black} Here, the full dynamics in \eqref{Eqn:TBH_eq} has 50-dimensional complex variables and the four conditional densities (in previous paragraph) are proposed as the closure model for the dynamics of the missing components, $\{u^2,\ldots, u^{\Lambda}\}$. Each of the four conditional densities above is a real-valued function that takes $m+n$ dimensional variables. In our numerical experiments, we will take $m+n \geq 20$.}

We compare three closure models of (\ref{Eqn:TBH_reduced}) with different
memory terms $m$ and $n$ and different temporal steps $\tau = 0.01$ or $0.1$. Figure \ref%
{Fig6_TBH} displays long-time statistics and short-time predictions for
these closure models. One can see from Figs.~\ref{Fig6_TBH}(a)(b)(c)(d)
that the long-time statistics can be well reproduced {\color{black}by the proposed closure models of (%
\ref{Eqn:TBH_reduced}) when the irrelevant variables have}
long memory terms, that is, $n$ is large enough. In terms of short-time predictions, all three closure models exhibit comparable results  for RMSEs and ANCRs where the errors saturate at about the time when the autocovariance function diminishes [Figs. \ref{Fig6_TBH}(e) and \ref{Fig6_TBH}(f)]. {\color{black} The fact that the two choices of $m, n, \tau$ ($m=n=20, \tau=0.1$ and $m=n=200, \tau=0.01$), corresponding to the two models with the same memory length $n\tau=2$ time units, produce comparable results (see the red dash-dotted and black dashed curves in Fig.~\ref{Fig6_TBH}) suggests that the temporal step $\tau$ does not affect the inference. Thus it is more economical to use the model with smaller $n$ (and possibly coarser time lag $\tau$) that gives the same accuracy.  Finally, we should also point out that if the memory length $n\tau$ (in unit time) is small, the estimates become less accurate.} Therefore, for this difficult test problem involving observations of the first Fourier mode of the TBH model, the proposed closure model can replicate the long-time statistics accurately and produce reasonable short-time prediction skills when there are long enough memory terms.

\section{Summary and discussion}\label{section5}

In this paper, we considered a data-driven nonparametric model for capturing the missing dynamics in the context of systems of ergodic SDE's and ODE's. The non-Markovian closure model is formulated as an averaging over an equilibrium conditional density function, $p(y_t|\bm{x}_{t-m:t},\bm{y}_{t-n:t-1})$, that is approximated using the kernel embedding of conditional distribution formulation. In particular, we considered a representation of the conditional density on RKHS induced by an orthonormal basis of appropriate weighted Hilbert space. A thorough investigation of the modeling framework on a linear Gaussian problem shows the consistency with the classical averaging theory for fast-slow systems and justifies our use of long non-Markovian memory terms to obtain accurate two-point statistical predictions in the case of no temporal scale separation. Numerical simulations on nonlinear problems demonstrate the robustness of the framework in producing accurate short-term predictions as well as to recover two-point statistics even when the missing terms are high-dimensional and have no separation of scales.

Modeling of missing dynamics with parametric models (or closure) has a long history as we noted in the introduction. Practically, such modeling paradigm requires modeler to choose the parametric model (ansatz) and fit the proposed model to the data to estimate the parameters in the ansatz. The choice of the parametric model is typically problem specific. When the underlying full system is known, one can deduce the model from the first principle. For example, one can apply the Mori-Zwanzig formalism to deduce such parametric model (see e.g., \cite{chk:02,chorin2007problem,hl:15,kondrashov2015data}) and then use various mathematical tools to estimate the memory integral terms as well as the parameters in the reduced model, which remains challenging if the resulting closure model is nonlinear or contains high-dimensional parameters. For example, when a rational approximation is used as a model for the memory kernel \cite{Lei14183}, while the parameters can be identified from derivatives of the kernel, it requires the availability of highly accurate time series (in the sense of accurate several order of derivatives) which is rare in practice.

In this paper, the proposed nonparametric formulation discovered some of the well-known parametric models, including the non-autonomous autoregressive linear models. An important feature of the proposed nonparametric framework in this paper is that it translates the problem of choosing parametric model into choosing the memory length $m,n$ and constructing orthonormal basis of a weighted Hilbert space of functions that take values on $\bm{z}\in\mathcal{Z}$. For the memory length, our experience suggests that we can use the decaying time scale of processes $x$ and $y$ as a guideline. While the natural candidate of model is a representation on a Hilbert space spanned by the orthonormal basis of functions that respect the geometry and sampling density of the data as in \cite{jiang2018parameter}, constructing such a basis is computationally challenging especially if $\mathcal{Z}$ is high-dimensional. In addition to the difficulty in the basis construction, the main computational cost arises as we evaluate the estimated basis functions on new points for future-time prediction. For very low-dimensional $\mathcal{Z}$, our numerical results suggest that we can avoid all of this practical issue with classical polynomial basis functions. In this case, the form of parametric model is polynomial functions. For very high-dimensional $\mathcal{Z}$, we showed the effectiveness of using the POD basis for representing linear problems. In nonlinear problems, we found that in some case, additional noise terms can be used to compensate for the orthogonal components that are not represented by the POD bases. In this case, the resulting parametric model is a linear non-autonomous autoregressive model. The second important feature is that the proposed nonparametric framework provides a linear technique for estimating the parameters in the resulting parametric models regardless of whether they are linear or nonlinear. This important feature is inherited from the kernel embedding formulation that allows one to ``gain'' linearity by representing nonlinear functions of a finite dimensional space $\mathcal{Z}$ with a basis of functions of infinite dimensional linear space. To summarize, the proposed framework and the results in this study suggest that one can understand the parametric modeling paradigm from a unified framework using appropriate Reproducing Kernel Hilbert Spaces. {\color{black} Such realization lies on the interpretation of the Mercer's type kernels in \eqref{mercer}. While the so-called kernel ``trick'' uses Mercer kernels to avoid the evaluation of inner product in feature space, our view point is to use the Mercer kernels to construct the parametric model of interest by an appropriate choice of finite number of basis functions.} Thus, this framework turns the problem of finding the right closure model into a problem of constructing a complete basis of the Hilbert space induced by the data, which remains challenging in general.

Finally, we should also point out that the modeling framework introduced here can also be realized with any supervised learning algorithm other than the kernel embedding discussed here. In a separate report, \cite{HJLY:19}, we found that the closure modeling framework introduced here is effective for high-dimensional nonlinear problems when it is realized with the Long-Short-Term-Memory (a special class of Recurrent Neural Network).

\section*{Acknowledgments}
It is a great pleasure to dedicate this paper to Andrew Majda on the occasion of his 70th birthday. The research of J.H. was partially supported by the ONR Grant N00014-16-1-2888, NSF Grants DMS-1619661 and DMS-1854299. S.W.J. was supported as a postdoctoral fellow under the ONR Grant N00014-16-1-2888.

\section*{Conflict of interest}

The authors declare that they have no conflict of interest.

\appendix

{\color{black}
\section{Kernel mean embedding of conditional distributions}\label{app:A}

The purpose of this review is to verify Eq.~\eqref{kme}. While the derivation here follows closely the description in \cite{Song2009hilbert,Song2013IEEE}, we taylor the discussion here for Mercer-type kernels induced by orthonormal basis of $L^2$-spaces. Some of the basic theory of RKHS can be found in many texts, such as \cite{christmann2008support}.

First, let us repeat the discussion in Section~\ref{sec21} on $\mathcal{Z}$. Let $\mathcal{Z}$ be a compact set and define $\hat{K}:\mathcal{Z}\times\mathcal{Z}\to\mathbb{R}$ to be a kernel, which means it is symmetric positive definite and let it be bounded. By Moore-Aronszajn theorem, there exists a unique Hilbert space $\mathcal{H}_Z=\overline{\mbox{span}\{\hat{K}(\bm{z},\cdot),\forall \bm{z}\in\mathcal{Z}\}}$. Let $\hat{q}:\mathcal{Z}\to\mathbb{R}$ be a positive weight function and $\{\varphi_k\}_{k\geq 1}$ be a set of eigenfunctions corresponding to eigenvalues $\{\xi_k\}$ of the following integral operator $\mathcal{\hat{K}}:L^2(\mathcal{Z},\hat{q}) \to L^2(\mathcal{Z},\hat{q})$, defined as,
\BEA
\mathcal{\hat{K}} f(\bm{z}) := \int_{\mathcal{Z}} \hat{K}(\bm{z},\bm{z}') f(\bm{z}') \hat{q}(\bm{z}') d\bm{z}'.\label{intop2}
\EEA
By Mercer's theorem, the kernel $\hat{K}$ has the following representation,
\BEA
\hat{K}(\bm{z},\bm{z}') = \sum_{k=1}^{\infty} \xi_k \varphi_k(\bm{z})\varphi_k(\bm{z}').\label{mercerkhat}
\EEA
We should point out that if $\mathcal{Z}$ is not a compact domain such as $\mathbb{R}^n$, with an exponentially decaying $\hat{q}$, one can construct a bounded Mercer-type kernel as in \eqref{mercerkhat} with an appropriate choice of decreasing sequence $\{\xi_k\}$ (see Lemma 3.2 in \cite{ZHL:19}) and it is a reproducing kernel corresponding to the RKHS $\mathcal{H}_Z$ (see Proposition 3.4 in \cite{ZHL:19}).

In this case, the RKHS $\mathcal{H}_Z$ induced by the Mercer-type kernel in \eqref{mercerkhat} is a subspace of $L^2(\mathcal{Z},\hat{q})$ with the reproducing property corresponding to an inner product defined as $\langle f,g\rangle_{\mathcal{H}_Z} = \sum_{k=1}^\infty \frac{f_k g_k}{\xi_k}$, for all $f,g\in\mathcal{H}_Z$ where $f_k = \langle f,\varphi_k\rangle_{L^2(\mathcal{Z},\hat{q})}$ and  $g_k = \langle g,\varphi_k\rangle_{L^2(\mathcal{Z},\hat{q})}$ . Then for any $f\in \mathcal{H}_Z$ and $\bm{z}\in\mathcal{Z}$, we can represent
\BEA
f(\bm{z}) = \langle f,\hat{K}(\bm{z},\cdot)\rangle_{\mathcal{H}_Z} = \sum_{k=1}^\infty \frac{f_k \xi_k \varphi_k(\bm{z})}{\xi_k} =  \sum_{k=1}^\infty f_k \varphi_k(\bm{z}),\label{expansion}
\EEA
with basis of $L^2(\mathcal{Z},\hat{q})$, where the convergence of the series holds uniformly (or in $C_0(\mathbb{R}^n)$ for non-compact $\mathcal{Z}=\mathbb{R}^n$).

We called the Hilbert space of functions, $\mathcal{H}_Z$, as an RKHS induced by the orthonormal basis of $L^2(\mathcal{Z},\hat{q})$. While we have discussed $\mathcal{H}$ as an RKHS induced by the orthonormal basis of $L^2(\mathcal{Y},q^{-1})$ in Section~\ref{sec21}, we can also repeat the argument above and construct $\mathcal{H}_Y$ as an RKHS induced by the orthonormal basis of $L^2(\mathcal{Y},q)$. In this case, recall that while $\{\psi_k q\}$ are orthogonal eigenbasis of the integral operator in \eqref{intoperator}, the orthogonal basis $\psi_k\in L^2(\mathcal{Y},q)$ are eigenfunctions of an adjoint integral operator of \eqref{intoperator}. That is, one can verify that
\BEA
\langle \psi_kq, \mathcal{K}^* \psi_k \rangle_{L^2(\mathcal{Y})} = \langle \mathcal{K}(\psi_kq), \psi_k\rangle_{L^2(\mathcal{Y})} = \lambda_k\langle \psi_kq,\psi_k\rangle_{L^2(\mathcal{Y})}, \label{adjoint}
\EEA
where for $f\in L^2(\mathcal{Y},q)$,
\BEA
\mathcal{K}^* f(x) := \int_\mathcal{Y} K^*(x,y) f(y) q(y)\,dy,\nonumber
\EEA
and $K^*(x,y)= q(x)^{-1}K(x,y)q^{-1}(y)$ is also a symmetric positive definite kernel. By Mercer's theorem, one can write
\BEA
K^*(y,y') = \sum_{k=1}^{\infty} \lambda_k \psi_k(y)\psi_k(y').\label{mercerkstar}
\EEA

Let $Y$ and $Z$ be random variables on $\mathcal{Y}$ and $\mathcal{Z}$ with distribution $P(Y,Z)$, we define the cross-covariance operators, $\mathcal{C}_{YZ}:\mathcal{H}_Z\to \mathcal{H}_Y$ and $\mathcal{C}_{ZZ}:\mathcal{H}_Z\to\mathcal{H}_Z$ as,
\BEA
\begin{aligned}\label{crosscov}
\mathcal{C}_{YZ} := \mathbb{E}_{YZ} [K^*(Y,\cdot)\otimes \hat{K}(Z,\cdot) ], \\
\mathcal{C}_{ZZ} := \mathbb{E}_{Z} [\hat{K}(Z,\cdot)\otimes \hat{K}(Z,\cdot) ].
\end{aligned}
\EEA
One can immediately see that for any  $f\in\mathcal{H}_Y$ and $g\in\mathcal{H}_Z$,
\BEA
\mathbb{E}_{YZ}[f(Y)\otimes g(Z)] &=& \int_{\mathcal{Y}\times\mathcal{Z}} f(y)g(\bm{z}) dP(y,\bm{z}) =  \int_{\mathcal{Y}\times\mathcal{Z}}  \langle f,K^*(y,\cdot)\rangle_{\mathcal{H}_Y} \langle g,\hat{K}(\bm{z},\cdot)\rangle_{\mathcal{H}_Z} dP(y,\bm{z}) \nonumber \\
&=& \int_{\mathcal{Y}\times\mathcal{Z}}  \langle f\otimes g, K^*(y,\cdot)\otimes \hat{K}(\bm{z},\cdot)\rangle_{\mathcal{H}_Y\otimes \mathcal{H}_Z} dP(y,\bm{z})  =  \langle f\otimes g, \mathcal{C}_{YZ}\rangle_{\mathcal{H}_Y\otimes \mathcal{H}_Z}. \label{jointexp}
\EEA
Let us define feature maps $\Psi:\mathcal{Y}\to\mathcal{F}_Y\subset \ell_2$ and $\Phi:\mathcal{Z}\to\mathcal{F}_Z\subset \ell_2$, respectively,
\BEA
\begin{aligned}\label{featuremaps}
\Psi(y) &= (\sqrt{\lambda_1}\psi_1(y), \sqrt{\lambda_2}\psi_2(y),\ldots), \\
\Phi(\bm{z}) &= (\sqrt{\xi_1}\varphi_1(\bm{z}), \sqrt{\xi_2}\varphi_2(\bm{z}),\ldots).
\end{aligned}
\EEA
Then we can write
\BEA
\hat{K}(\bm{z},\bm{z}') &=& \langle \Phi(\bm{z}),\Phi(\bm{z}')\rangle_{\ell_2} = \langle \hat{K}(\bm{z},\cdot),\hat{K}(\bm{z}',\cdot)\rangle_{\mathcal{H}_Z}, \nonumber\\
K^*(y,y') &=& \langle \Psi(y),\Psi(y')\rangle_{\ell_2} = \langle K^*(y,\cdot),K^*(y',\cdot)\rangle_{\mathcal{H}_Y}, \nonumber
\EEA
where the inner products in $\mathcal{H}_Z$ and $\mathcal{H}_Y$ can be identified by $\ell_2$ inner products in the corresponding feature spaces.  Also, for any function $f\in\mathcal{H}_Z$ and $\bm{z}\in\mathcal{Z}$,  we can rewrite the expansion in \eqref{expansion} as,
\BEA
f(\bm{z}) =  \langle f,\hat{K}(\bm{z},\cdot) \rangle_{\mathcal{H}_Z} = \sum_{k=1}^\infty  \langle f,\varphi_k \rangle_{L^2(\mathcal{Z},\hat{q})}  \varphi_k(\bm{z}) =  \sum_{k=1}^\infty  \frac{\langle f,\varphi_k \rangle_{L^2(\mathcal{Z},\hat{q})}}{\sqrt{\xi_k}}  \Phi_k(\bm{z}) = \sum_{k=1}^\infty\langle f, \Phi_k\rangle_{\mathcal{H}_Z} \Phi_k(\bm{z}), \label{rkhsexpansion}
\EEA
where we have defined the functions $\Phi_k = \sqrt{\xi_k}\varphi_k \in \mathcal{H}_Z$. For convenience of the discussion below, we also define the functions $\Psi_k:=\sqrt{\lambda_k}\psi_k\in\mathcal{H}_Y$.

Using the identity in \eqref{jointexp}, we can represent the cross-operators in  \eqref{crosscov} on the basis coordinates $\Psi_k \in \mathcal{H}_Y$ and $\Phi_\ell \in \mathcal{H}_Z$ as follows:
\BEA
\begin{aligned}\label{matrixdefn}
\left[C_{YZ}\right]_{k\ell}&:=\mathbb{E}_{YZ}[\Psi_k(Y)\otimes \Phi_\ell(Z)] = \langle \Psi_k\otimes \Phi_\ell, \mathcal{C}_{YZ}\rangle_{\mathcal{H}_Y\otimes \mathcal{H}_Z},\\
\left[C_{ZZ}\right]_{k\ell}&:=\mathbb{E}_{ZZ}[\Phi_k(Z)\otimes \Phi_\ell(Z)]=\langle \Phi_k\otimes \Phi_\ell, \mathcal{C}_{ZZ}\rangle_{\mathcal{H}_Z\otimes \mathcal{H}_Z}=\langle \Phi_k, \mathcal{C}_{ZZ} \Phi_\ell\rangle_{\mathcal{H}_Z} .
\end{aligned}
\EEA

Thus, the components of the following matrix multiplication are given as,
\BEA
\left[C_{YZ}C_{ZZ}^{-1} \right]_{k\ell} &=& \sum_{j} \left[C_{YZ}\right]_{kj}\left[C_{ZZ}^{-1}\right]_{j\ell} \nonumber\\
&=& \sum_j \langle \Psi_k\otimes \Phi_j, \mathcal{C}_{YZ}\rangle_{\mathcal{H}_Y\otimes \mathcal{H}_Z}\langle \Phi_j, \mathcal{C}_{ZZ}^{-1}\Phi_\ell\rangle_{\mathcal{H}_Z} \nonumber \\
&=&\left\langle \mathcal{C}_{YZ},  \Psi_k\otimes \left( \sum_j \langle \Phi_j, \mathcal{C}_{ZZ}^{-1}\Phi_\ell \rangle_{\mathcal{H_Z}} \Phi_j  \right)  \right\rangle_{\mathcal{H}_Y\otimes \mathcal{H_Z}} \nonumber \\
&=&\left\langle \mathcal{C}_{YZ},  \Psi_k \otimes \mathcal{C}_{ZZ}^{-1} \Phi_\ell   \right\rangle_{\mathcal{H}_Y\otimes \mathcal{H}_Z}  \nonumber\\
&=&\left\langle \mathcal{C}_{YZ} \mathcal{C}_{ZZ}^{-1},  \Psi_k \otimes \Phi_\ell   \right\rangle_{\mathcal{H}_Y\otimes \mathcal{H}_Z} \nonumber \\
&=&\left\langle \mathcal{C}_{YZ} \mathcal{C}_{ZZ}^{-1}\Phi_\ell,  \Psi_k    \right\rangle_{\mathcal{H}_Y}.\label{matrixmultiplication}
\EEA
To clarify this derivation, the second equality used the definition in \eqref{matrixdefn}, the fourth line used the fact that $\mathcal{C}_{ZZ}^{-1}\Psi_\ell\in \mathcal{H}_Z$ can be expanded as in \eqref{rkhsexpansion}, and the rest of the lines used the standard tensor identity.

The theory of kernel mean embedding of conditional distributions (see \cite{Song2009hilbert,Song2013IEEE}) suggests that,
\BEA
\mathbb{E}_{Y|\bm{z}} [\Psi_k(Y)] = \langle \Psi_k,  \mathcal{C}_{YZ} \mathcal{C}_{ZZ}^{-1} \hat{K}(\bm{z},\cdot) \rangle_{\mathcal{H}_Y}.\label{song}
\EEA
Since $\hat{K}(\bm{z},\cdot) \in \mathcal{H}_Z$, we can employ the expansion in \eqref{rkhsexpansion} and deduce,
\BEA
\mathbb{E}_{Y|\bm{z}} [\Psi_k(Y)]
&=& \langle \Psi_k,  \mathcal{C}_{YZ} \mathcal{C}_{ZZ}^{-1} \sum_{j=1}^{\infty} \frac{\langle  \hat{K}(\bm{z},\cdot),\varphi_j\rangle_{L^2(\mathcal{Z},\hat{q})}}{\sqrt{\xi_j}}\Phi_j \rangle_{\mathcal{H}_Y} \nonumber \\
&=& \sum_{j=1}^{\infty} \frac{\langle  \hat{K}(\bm{z},\cdot),\varphi_j\rangle_{L^2(\mathcal{Z},\hat{q})}}{\sqrt{\xi_j}} \langle \Psi_k,  \mathcal{C}_{YZ} \mathcal{C}_{ZZ}^{-1} \Phi_j \rangle_{\mathcal{H}_Y} \nonumber \\
&=&  \sum_{j=1}^{\infty}   \frac{1}{\sqrt{\xi_j}}\left[C_{YZ}C_{ZZ}^{-1} \right]_{kj} \int _{\mathcal{Z}} \hat{K}(\bm{z},\bm{z}')  \varphi_j(\bm{z}')\hat{q}(\bm{z}')\,d\bm{z}'   \nonumber \\
&=& \sum_{j=1}^{\infty}  \left[C_{YZ}C_{ZZ}^{-1} \right]_{kj} \Phi_j(\bm{z}),  \label{kmde}
\EEA
where we have used \eqref{matrixmultiplication} to deduce the third equality above and used the fact that $\varphi_j$ and $\xi_j$ are eigenfunction and eigenvalue of the integral operator in \eqref{intop2}.
Define,
\BEA
\left[\bm{C}_{{YZ}}\right] _{ks} =\mathbb{E}_{{YZ}}%
\left[ \psi _{k}(Y)\otimes \varphi _{s}(Z)\right], \quad\quad
\left[ \bm{C}_{{ZZ}}\right] _{sl} =\mathbb{E}_{{ZZ}}%
\left[ \varphi _{s}(Z)\otimes\varphi _{l}(Z)\right],\nonumber
\EEA
then from \eqref{matrixdefn} and the definitions of the corresponding feature maps in \eqref{featuremaps},
\BEA
\left[C_{{YZ}}\right] _{ks}  = \sqrt{\lambda_k\xi_s}\left[\bm{C}_{{YZ}}\right] _{ks}, \quad\quad
\left[C_{{ZZ}}\right] _{sl}  = \sqrt{\xi_s\xi_l}\left[\bm{C}_{{ZZ}}\right] _{sl}, \quad\quad \left[C_{YZ}C_{ZZ}^{-1} \right]_{k\ell} = \frac{\sqrt{\lambda_k}}{\sqrt{\xi_l}} \left[\bm{C}_{YZ}\bm{C}_{ZZ}^{-1} \right]_{k\ell}.\nonumber
\EEA
Substituting the third equation above to \eqref{kmde} and using the definitions of the feature maps in \eqref{featuremaps}, we obtain
\BEA
\mathbb{E}_{Y|\bm{z}} [\psi_k(Y)] = \frac{1}{\sqrt{\lambda_k}} \sum_{j=1}^{\infty}  \left[C_{YZ}C_{ZZ}^{-1} \right]_{kj} \Phi_j(\bm{z}) = \sum_{j=1}^{\infty}  \left[\bm{C}_{YZ}\bm{C}_{ZZ}^{-1} \right]_{kj} \varphi_j(\bm{z}),\nonumber
\EEA
which is exactly the claim in \eqref{kme}.
}

\section{\label{app:ACF_linear}ACV of the  multi-scale linear Gaussian model}

The full model (\ref{Eqn:eqn_x})-(\ref{Eqn:eqn_y}) can be rewritten as
\begin{eqnarray}
\dot{x} &=&\left( a_{11}x+a_{12}y\right) +\sigma _{x}\xi _{x},
\label{Eqn:linear_Gauss} \\
\dot{y} &=&\frac{1}{\epsilon }\left( a_{21}x+a_{22}y\right) +\frac{\sigma
_{y}}{\sqrt{\epsilon }}\xi _{y},  \notag
\end{eqnarray}%
where $\xi _{x}$ and $\xi _{y}$\ are independent standard Gaussian noises.
Similarly, the closure model (\ref{modelwithmemory}) can be rewritten as
\begin{equation}
\dot{x}_t=\left( a_{11}x_t+a_{12}\Sigma _{12}\Sigma _{22}^{-1}\bm{x}\right)
+\sigma _{x}\xi _{x},  \label{Eqn:linear_rkhs_1}
\end{equation}%
where $\bm{x}:=\bm{x}_{t-m:t} = \left[ x_{t-m},x_{t-m+1},\ldots ,x_{t}\right] ^\top$ and $%
\Sigma _{12}$ and $\Sigma _{22}$ are defined in Eq. (\ref{Eqn:linear_SIG}).
To simplify the notation, we drop the time indices $t-m:t$. We also drop the ``hat"-notation
in $x_t$ and $\bm{x}_t$ since we will use it to denote the Fourier coefficient in this section.
In this Appendix, we prove that the autocovariance (ACV) function of the closure model (\ref%
{Eqn:linear_rkhs_1}) is approximately equal to that of the full model (\ref%
{Eqn:linear_Gauss}) for any value of $\epsilon$.

The Fourier transform and inverse Fourier transform is defined as%
\begin{equation*}
\widehat{f}\left( \omega \right) =\int f\left( t\right) e^{-i\omega t}dt,%
\text{ \ \ }f\left( t\right) =\frac{1}{2\pi }\int \widehat{f}\left( \omega
\right) e^{i\omega t}d\omega .
\end{equation*}%
The Fourier transforms of variables $x$ and $y$ of the full model (\ref%
{Eqn:linear_Gauss}) can be obtained as%
\begin{eqnarray}
\widehat{x} &=&\frac{\left( i\omega -\frac{1}{\epsilon }a_{22}\right) \sigma
_{x}\widehat{\xi }_{x}+a_{12}\frac{\sigma _{y}}{\sqrt{\epsilon }}\widehat{%
\xi }_{y}}{\left( i\omega -a_{11}\right) \left( i\omega -\frac{1}{\epsilon }%
a_{22}\right) -a_{12}\frac{1}{\epsilon }a_{21}},  \label{Eqn:xhat} \\
\widehat{y} &=&\frac{\left( i\omega -a_{11}\right) \frac{\sigma _{y}}{\sqrt{%
\epsilon }}\widehat{\xi }_{y}+\frac{1}{\epsilon }a_{21}\sigma _{x}\widehat{%
\xi }_{x}}{\left( i\omega -a_{11}\right) \left( i\omega -\frac{1}{\epsilon }%
a_{22}\right) -a_{12}\frac{1}{\epsilon }a_{21}}.  \label{Eqn:yhat}
\end{eqnarray}%
Then, for the full model (\ref{Eqn:linear_Gauss}), the resulting spectrum of
$x$ is%
\begin{equation*}
\left\vert \widehat{x}\left( \omega \right) \right\vert ^{2}=\frac{\left(
\omega ^{2}+c_{0}^{2}\right) \sigma _{x}^{2}\left\vert \widehat{\xi }%
_{x}\right\vert ^{2}+d_{0}^{2}\sigma _{y}^{2}\left\vert \widehat{\xi }%
_{y}\right\vert ^{2}}{\left( -\omega ^{2}+\omega _{0}^{2}\right) ^{2}+\gamma
_{0}^{2}\omega ^{2}},
\end{equation*}%
where
\begin{eqnarray*}
c_{0} &=&\frac{a_{22}}{\epsilon },\text{ \ \ }d_{0}=\frac{a_{12}}{\sqrt{%
\epsilon }},\text{ \ \ }\omega _{0}=\sqrt{\frac{1}{\epsilon }\left(
a_{11}a_{22}-a_{12}a_{21}\right) }, \\
\gamma _{0} &=&a_{11}+\frac{1}{\epsilon }a_{22},\text{ \ \ }\left\vert
\widehat{\xi }_{x}\right\vert ^{2}=1,\text{ \ \ }\left\vert \widehat{\xi }%
_{y}\right\vert ^{2}=1.
\end{eqnarray*}

Now we compute the Fourier transform of the closure model (\ref%
{Eqn:linear_rkhs_1}),%
\begin{equation}
i\omega \widehat{X}=a_{11}\widehat{X}+a_{12}\Sigma _{12}\Sigma _{22}^{-1}%
\left[
\begin{array}{c}
1 \\
e^{-i\omega \tau} \\
\vdots \\
e^{-i\omega m\tau}%
\end{array}%
\right] \widehat{X}+\sigma _{x}\widehat{\xi }_{x},  \label{Eqn:linear_four}
\end{equation}%
where $\widehat{X}$ is the Fourier transform of $x_t$ in Eq. (\ref{Eqn:linear_rkhs_1}).
We need to simplify the quantity \\
$\Sigma _{12}\Sigma _{22}^{-1}\left[
\begin{array}{cccc}
1 & e^{-i\omega \tau} & \cdots & e^{-i\omega m\tau}%
\end{array}%
\right] ^\top$ in Eq. (\ref{Eqn:linear_four}). Let $S=\Sigma _{12}\Sigma
_{22}^{-1}$ be the $1\times \left( m+1\right) $ vector with components denoted by $S\left[ n\right]$ for $n=0,\ldots ,m$. Then, we can write
\begin{equation}
\Sigma _{12}\Sigma _{22}^{-1}\left[
\begin{array}{c}
1 \\
e^{-i\omega \tau} \\
\vdots \\
e^{-i\omega m\tau}%
\end{array}%
\right] =S\left[
\begin{array}{c}
1 \\
e^{-i\omega \tau} \\
\vdots \\
e^{-i\omega m\tau}%
\end{array}%
\right] =\sum_{n=0}^{m}S\left[ n\right] e^{-i\omega n\tau} :=
\widehat{S}_m\left( \omega \right) ,  \label{Eqn:simp_conv}
\end{equation}%
which is nothing but the discrete Fourier transform of $S$. Notice that, for any $n=0,\ldots ,m$,
\BEA
\sum_{k=0}^m S\left[ k\right] \gamma_{xx,m}\left[n-k\right] =
\sum_{k=0}^m S\left[ k\right] \Sigma_{22}\left[k, n\right]  = \Sigma_{12}\left[ n\right]  = \gamma_{xy,m} \left[ n\right]
\EEA
where the first equality is due to the fact that the process is stationary such that $\Sigma_{22}[k,n] = \gamma_{xx,m}[n-k]$, the second equality is due to $S\Sigma_{22}=\Sigma_{12}$, and the last equality is by the definition of the covariance function. By the discrete convolution theorem, we have
\BEA
\widehat{S}_m\left( \omega \right) \widehat{\gamma }_{xx,m}\left( \omega \right) =\widehat{\gamma }_{xy,m} \left( \omega \right),  \label{Eqn:Sxy_conv}
\EEA
where $\widehat{\gamma }_{xx,m}$ and $\widehat{\gamma }_{xy,m}$ are the discrete Fourier
transforms of $\gamma _{xx,m}$ and $\gamma _{xy,m}$, respectively. Substituting $%
\widehat{S}_m\left( \omega \right) $ in Eq. (\ref{Eqn:Sxy_conv}) into Eq. (\ref%
{Eqn:simp_conv}), we obtain%
\begin{equation}
\Sigma _{12}\Sigma _{22}^{-1}\left[
\begin{array}{c}
1 \\
e^{-i\omega \delta t} \\
\vdots \\
e^{-i\omega m\delta t}%
\end{array}%
\right] = \frac{\widehat{\gamma }_{xy,m}\left( \omega \right) }{\widehat{%
\gamma }_{xx,m}\left( \omega \right) } \longrightarrow \frac{\widehat{\gamma }_{xy}\left( \omega \right) }{\widehat{\gamma }_{xy}\left( \omega \right) }, \quad\quad \mbox{as }m\to\infty,  \label{Eqn:apprx1}
\end{equation}%
where $\widehat{\gamma }_{xx}$ and $\widehat{\gamma }_{xy}$ denote the Fourier transform of the covariance functions $\gamma _{xx}$ and $\gamma _{xy}$.

Substituting the limiting case of Eq. (\ref{Eqn:apprx1}) into Eq. (\ref{Eqn:linear_four}%
), we can simplify the Fourier transform of the closure model as follows,%
\begin{equation}
i\omega \widehat{X}=a_{11}\widehat{X}+a_{12}\frac{\widehat{\gamma }%
_{xy}\left( \omega \right) }{\widehat{\gamma }_{xx}\left( \omega \right) }%
\widehat{X}+\sigma _{x}\widehat{\xi }_{x}.  \label{Eqn:reduced_simple}
\end{equation}%
Moreover, based on the Wiener-Khinchin theorem and the cross-correlation
theorem, we can further simplify Eq. (\ref{Eqn:reduced_simple}) as%
\begin{equation}
i\omega \widehat{X}=a_{11}\widehat{X}+a_{12}\frac{\widehat{y}}{\widehat{x}}%
\widehat{X}+\sigma _{x}\widehat{\xi }_{x}.  \label{Eqn:reduced_sim2}
\end{equation}%
Substituting Eqs. (\ref{Eqn:xhat}) and (\ref{Eqn:yhat}) into above Eq. (\ref%
{Eqn:reduced_sim2}), we obtain the Fourier transform of the relevant variable, $\widehat{X}$, of the closure model,%
\begin{equation}
\widehat{X}=\frac{\left( i\omega -\frac{1}{\epsilon }a_{22}\right) \sigma
_{x}\widehat{\xi }_{x}+a_{12}\frac{\sigma _{y}}{\sqrt{\epsilon }}\widehat{%
\xi }_{y}}{\left( i\omega -a_{11}\right) \left( i\omega -\frac{1}{\epsilon }%
a_{22}\right) -a_{12}\frac{1}{\epsilon }a_{21}},  \label{Eqn:Xhat}
\end{equation}%
which is the same as the $\widehat{x}$ of the full model in Eq. (\ref{Eqn:xhat}).
Therefore, the ACV of the closure model (\ref{Eqn:linear_rkhs_1})
is consistent with that of the full model (\ref{Eqn:linear_Gauss}) in the limit of $m\to \infty$.
In the numerics, the error comes from the truncation of finite number of memory terms
in Eq. (\ref{Eqn:simp_conv}).

\bibliographystyle{abbrv}
%\bibliography{Short_Format,Data_Analysis2}

\end{document}